\documentclass{aa}  

\usepackage{graphicx}
\usepackage{txfonts}
\usepackage{hyperref}
\begin{document}

   \title{Central galaxy alignments} 

   \subtitle{Dependence on the mass and the large-scale environment}

 \author{Facundo Rodriguez\thanks{facundo.rodriguez@unc.edu.ar}
   \inst{1,2}
          \and
          Manuel Merchán\inst{1,2}
          \and
          Daniela Galárraga-Espinosa\inst{3,4}
          \and
          Agustina V. Marsengo-Colazo\inst{1,2}
          \and
          Antonio D. Montero-Dorta\inst{5}
          \and
          Vicente Izzo Dominguez\inst{5}
          \and
          M. Celeste Artale\inst{6}
          }

   \institute{CONICET. Instituto de Astronomía Teórica y Experimental (IATE). Laprida 854, Córdoba X5000BGR, Argentina.
   \and
             Universidad Nacional de Córdoba (UNC). Observatorio Astronómico de Córdoba (OAC). Laprida 854, Córdoba X5000BGR, Argentina.
        \and
             Kavli IPMU (WPI), UTIAS, The University of Tokyo, Kashiwa, Chiba 277-8583, Japan
        \and
        Max Planck Institute for Astrophysics, Karl-Schwarzschild-Straße 1, D-85748 Garching, Germany.
        \and
             Departamento de F\'isica, Universidad T\'ecnica Federico Santa Mar\'ia, Avenida Vicu\~na Mackenna 3939, San Joaqu\'in, Santiago, Chile.
        \and
        Universidad Andres Bello, Facultad de Ciencias Exactas, Departamento de Ciencias Físicas, Instituto de Astrofísica, Av. Fernández Concha 700, Santiago, Chile.
             }

   \date{Received September 15, 1996; accepted March 16, 1997}

 
  \abstract
   { Observations indicate that central galaxies show a significant alignment of their main shape axes with other galaxies in their group, as well as with the large-scale structure of the universe. Simulations have corroborated this finding, providing further insights into how the shape of the stellar component aligns with the surrounding dark matter halo. Recent studies have also investigated the evolution of this alignment in bright central galaxies, revealing that the shapes of the dark matter halo and the stellar component can differ. These results suggest that assembly and merger processes have played a crucial role in the evolution of this alignment.}
   {In this work, we  aim at gaining a deeper understanding of galaxy alignments by quantifying how this property is related to the mass of the haloes hosting central galaxies and to the large-scale environment measured at different scales.}
   {By studying different angles, we describe how the alignments of central galaxies depend on the mass of the haloes they inhabit. We explore how the main axes of central galaxies align across different scales, both in three-dimensional and two-dimensional projections. We examine how halo mass influences these alignments and how they vary in the surrounding large-scale environment. Additionally, we analyze the characteristics of these alignments across different environments within the large-scale structure of the universe. To conduct this study, we employ the TNG300 hydrodynamical simulations and compare our results with the spectroscopic data from the Sloan Digital Sky Survey Data Release 18 (SDSS DR18).}
   {Three types of alignment were analysed: between stellar and dark matter components, between satellite galaxies and the central galaxy, and between the central galaxy and its host halo. The results show that the alignment increases with halo mass and varies with the environment (clusters, filaments, cluster periphery, and others). However, after controlling for local density, we found that most of the observed trends disappear, except for a marginal influence of cosmic filaments on some of the considered alignment angles. The SDSS observations confirm a mass dependence similar to the simulations, although observational biases limit the detection of differences between the different environments.}
   {}

    \keywords{large-scale structure of Universe -- Methods: statistical 
   -- Galaxies: halos -- dark matter -- Galaxies: groups: general 
               }

   \maketitle
%

\section{Introduction}

Galaxies form and evolve within a large-scale structure predominantly made up of dark matter (DM). Their spatial distribution is influenced by gravitational forces acting initially on tiny fluctuations in the early universe. As structures develop hierarchically through gravitational instability, the tidal fields, the process of matter accretion, and even baryonic physics are expected to subtly affect the properties of galaxies and DM halos. In this scenario, galaxies will tend to have preferred shapes, orientations, and distributions depending on the halo in which they form and, potentially, the surrounding local and large-scale environment. 

Several observational works have investigated these so-called {\emph{intrinsic alignments}}, which are expected to have important implications for the extraction of cosmological information from upcoming galaxy surveys (e.g. \citep{Hirata2007,Troxel2015,Joachimi2015,Kirk2015}. Evidence suggests that the alignment of galaxies with each other and with large-scale cosmic structures is influenced not only by their luminosity, colour and star formation history, but also their position within the host halo and cosmic environment. In this context, red/elliptical galaxies in general tend to display more significant alignments on a variety of scales \citep[see e.g.][]{Sales2004,yang2006alignment,Agustsson2010,Kirk2015, Rodriguez2022, Smith2023, Desai2022}. Moreover, red satellite galaxies show a stronger preference to align with the galactic plane of red central galaxies \citep{Yang2005,Wang2008,Kiessling2015,Kirk2015,Libeskind2015,Welker2018,Pawlowski2018,Johnston2019}. This specific connection between star formation activity, or lack thereof, and galaxy alignments is sometimes referred to as \emph{anisotropic quenching} or \emph{angular conformity} \citep{Wang2008,Stott2022,Ando2023}, an obvious reference to the correlation effect called galactic conformity \citep[see][and references therein]{Bray2016,Otter2020,Maier2022,lacerna2022}.

The shape of spiral galaxies, on the other hand, is usually linked to their angular momentum,  which has been proposed to arise from torques produced by the external gravitational field \citep{Heavens2000, Catelan2001, Codis2015}, and tends to align parallel to filaments, in contrast to the angular momentum of higher-mass galaxies, which is preferentially perpendicular (\cite{Welker2020, Barsanti2022, Kraljic2021}, and e.g. \cite{Ganeshaiah-Veena2019, Ganeshaiah-Veena2021} for studies in simulations). Despite this connection, there is little observational evidence for shape alignments in spiral galaxies 
\citep{Zjupa2020, Johnston2019, Samuroff2023}. However, some studies based on observations and simulations indicate that central disk galaxies show greater misalignment with their host haloes, and that this misalignment decreases with increasing halo mass and ex situ stellar mass fraction \citep{Xu2023_SDSS,Xu2023_TNG300}.

In this context, a useful tool to assess the degree of alignment of a galaxy or group population with respect to any reference system is provided by the anisotropic correlation function, which is used to compare the clustering of different sub-regions with respect to a given orientation axis \citep{paz2008angular,paz2011alignments}. \cite{Rodriguez2022} applied this method to the spectroscopic data provided by Sloan Digital Sky Survey Data Release 16 \citep[SDSS DR16,][]{ahumada2020}, in combination with a group identification scheme \citep{rodriguez20}, to show that bright central galaxies align with both the satellites inhabiting the same halo and the nearby large scale cosmic structures up to a distance of $\gtrsim$ 10 Mpc. They also reported a significant dependence on colour, with red central galaxies being more aligned with their environment than the blue
ones. A physical interpretation of these observational constraints was subsequently provided in \cite{Rodriguez2023} based on a detailed study using the  IllustrisTNG\footnote{\url{http://www.tng-project.org}} (hereafter simply TNG) hydrodynamical simulation. The intrinsic alignments of \cite{Rodriguez2023} are decomposed in a series of correlations and subdependences across scales, where the alignment between the baryonic component of the central galaxy and the large-scale structure is mediated by internal alignments with the subhalo and the host halo. 

The results of \cite{Rodriguez2022, Rodriguez2023} prompted the additional analysis of \cite{Rodriguez2024}, where the evolution of the intrinsic alignments was investigated for red and blue galaxies separately, in connection with their different assembly and merger histories. Mergers are in fact shown to be a major contributor to the buildup of these alignments for red central galaxies, as opposed to blue central galaxies, which are younger, experience fewer mergers, and have only recently acquired their oblate shape. This aligns with findings from other cosmological simulations, such as those in \cite{Lagos2018}, based on the EAGLE simulations, which demonstrate the broader prevalence of merger-induced alignment effects. The work of \cite{Rodriguez2024} emphasizes even more the multi-scale nature of the intrinsic alignments. Mergers tend to increase the alignment between the stellar content of central galaxies and their dark-matter halos, whereas the preferred orientation of halos with respect to the large-scale environment is initially stronger and tends to diminish with time.

In this work, we use both observational data from the SDSS and simulation data from TNG to investigate in greater detail the dependence of central galaxy alignments on the environment. Our analysis follows two approaches. First, we examine the dependence on stellar and halo mass, which are known to be first-order proxies for local and large-scale environments. Second, we study how alignments vary with the location of galaxies within the cosmic web. The different components of the cosmic web are identified using the Discrete Persistent Structures Extractor \citep[DisPerSE;][]{Sousbie2011a,Sousbie2011b}, a structure finder that detects critical points of the density field (maxima, minima and saddle points). By combining the DisPerSE output with a halo/group catalogue, galaxies can be classified into specific cosmic environments, such as filaments, clusters, or cluster outskirts.
This analysis complements several other works where DisPerSE has been employed to investigate the link between the properties of galaxies and halos and the cosmic web, including the dependence of gas accretion and star formation in galaxies \citep{galarraga2023}, galaxy clustering and assembly bias \citep{MonteroDorta2024}, halo occupation distributions \citep{Perez2024A,Perez2024B}, or occupancy variations \citep{Wang2024}, to name but a few.   

The paper is organized as follows. Section \ref{data} describes the simulation (TNG) and observational (SDSS) data employed in this work, along with the identification of cosmic-web environments in both data sets based on DisPerSE. The different alignments measured in this work are defined in Section \ref{sec:alignments}. The dependence of the alignments on both mass and cosmic-web environment as measured from TNG is presented in Sections \ref{sec:mass} and \ref{sec:environment}, respectively. Section \ref{sec:sdss} focuses on the measurements performed on the SDSS, including a comparison with the simulation-based results. Finally, Section \ref{sec:conclusions} summarizes the main conclusions of our work, while discussing their implications and potential ramifications. The TNG simulation adopts the standard $\Lambda$CDM cosmology \citep{Planck2016}, with parameters $\Omega_{\rm m} = 0.3089$,  $\Omega_{\rm b} = 0.0486$, $\Omega_\Lambda = 0.6911$, $H_0 = 100\,h\, {\rm km\, s^{-1}Mpc^{-1}}$ with $h=0.6774$, $\sigma_8 = 0.8159$, and $n_s = 0.9667$.


\section{Data}
\label{data}

\subsection{The TNG hydrodynamical simulation} 

In this study, we utilize the galaxy and dark-matter halo catalogues from the Illustris TNG300 simulation at ( $z=0$ ) (referred to as TNG300, \citealt{Nelson2019, Pillepich2018}). The TNG magneto-hydrodynamical cosmological simulations are executed using the {\sc arepo} moving-mesh code \citep{Springel2010} and represent an enhanced iteration of the original Illustris simulations \citep{Vogelsberger2014a, Vogelsberger2014b, Genel2014}. These simulations incorporate sub-grid models that address factors such as radiative metal-line gas cooling, star formation, chemical enrichment from SNII, SNIa, and AGB stars, as well as stellar feedback processes, formation of supermassive black holes with multi-mode quasar activity, and kinetic feedback from black holes. It adopts a cubic box with a side length of $205\,h^{-1}$~Mpc and is run with 2500$^3$ dark-matter particles, each with a mass of $4.0 \times 10^7 h^{-1} {\rm M_{\odot}}$. The initial conditions include 2500$^3$ gas cells, with each gas cell having a mass of $7.6 \times 10^6 h^{-1} {\rm M_{\odot}}$.

A friends-of-friends (FOF) algorithm is employed to identify dark matter haloes—referred to as groups in this paper—using a linking length of 0.2 times the mean inter-particle separation \citep{Davis1985}. The gravitationally bound substructures, hereafter referred to as subhaloes, are identified using the SUBFIND algorithm \citep{Springel2001, Dolag2009}. 
We define the stellar mass of galaxies, denoted as ${\rm M_\ast{}}$, as the total mass of all stellar particles associated with each subhalo. In our analysis, following the selection criteria used in many previous works that take simulation resolution into account \citep[e.g.][]{Pillepich2018b,Rodriguez2023,Perez2024,Rodriguez2024}, we consider all simulated galaxies with stellar mass ($M_\star$) greater than $10^{8.5}$~M$_{\odot}$, resulting on a total number of 429982 galaxies. 
In this study, we analyze central galaxies identified as those subhalos whose index matches the \texttt{GroupFirstSub} entry of their corresponding FoF group in the TNG300 simulation, selecting those with $M_r < -19.5$ in the $r$-band. This yields a final sample of 89,534 systems,  each containing more than 880 dark matter particles and 100 stellar particles. The magnitude threshold matches the group-definition criterion from \citet{rodriguez2020}, where groups must contain at least one galaxy brighter than $M_r = -19.5$ (defining the faintest observable centrals in optical surveys).
 
\subsection{SDSS}
\label{datasdss}
\begin{figure}
    \centering
    \includegraphics[width=0.9\columnwidth]{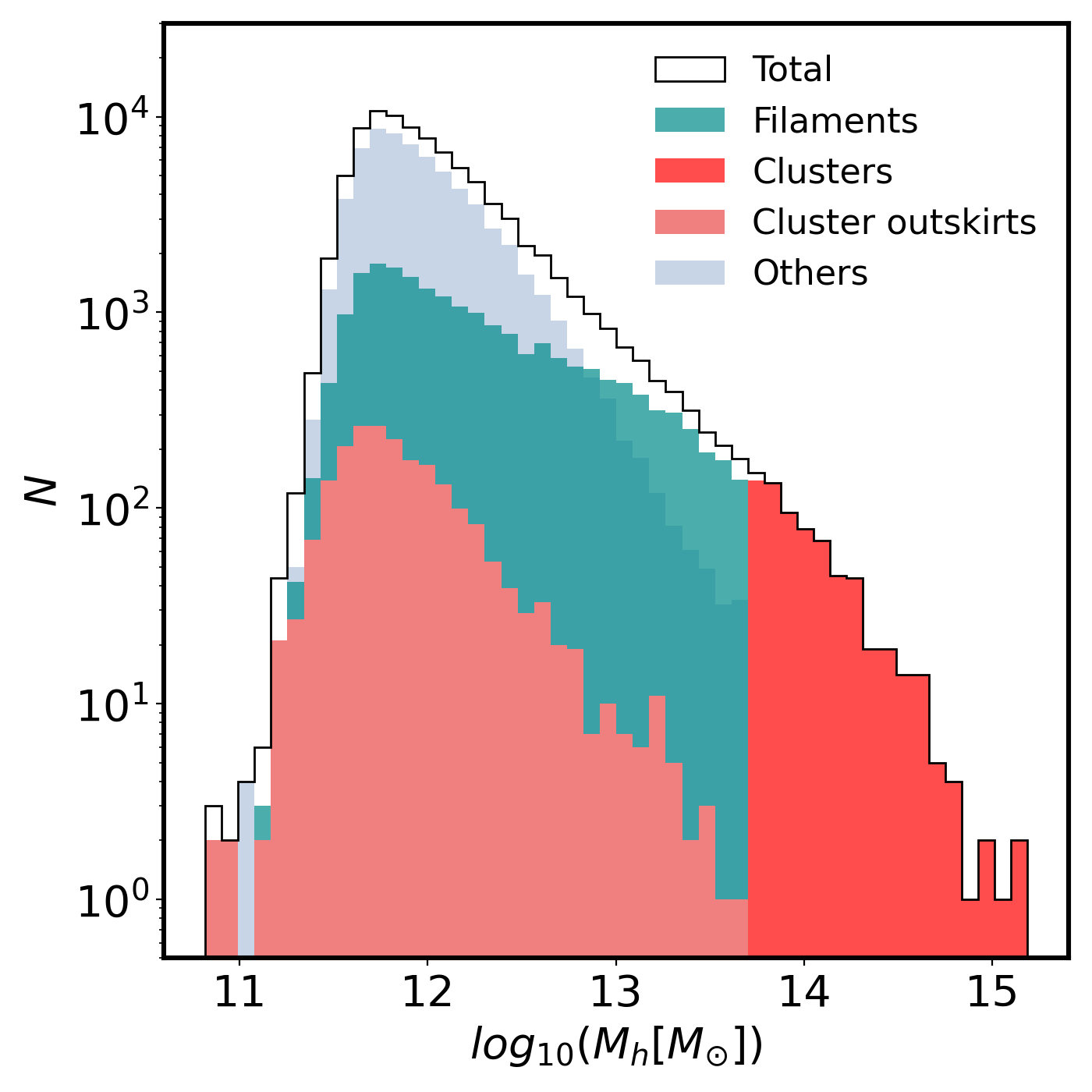}
    \includegraphics[width=0.9\columnwidth]{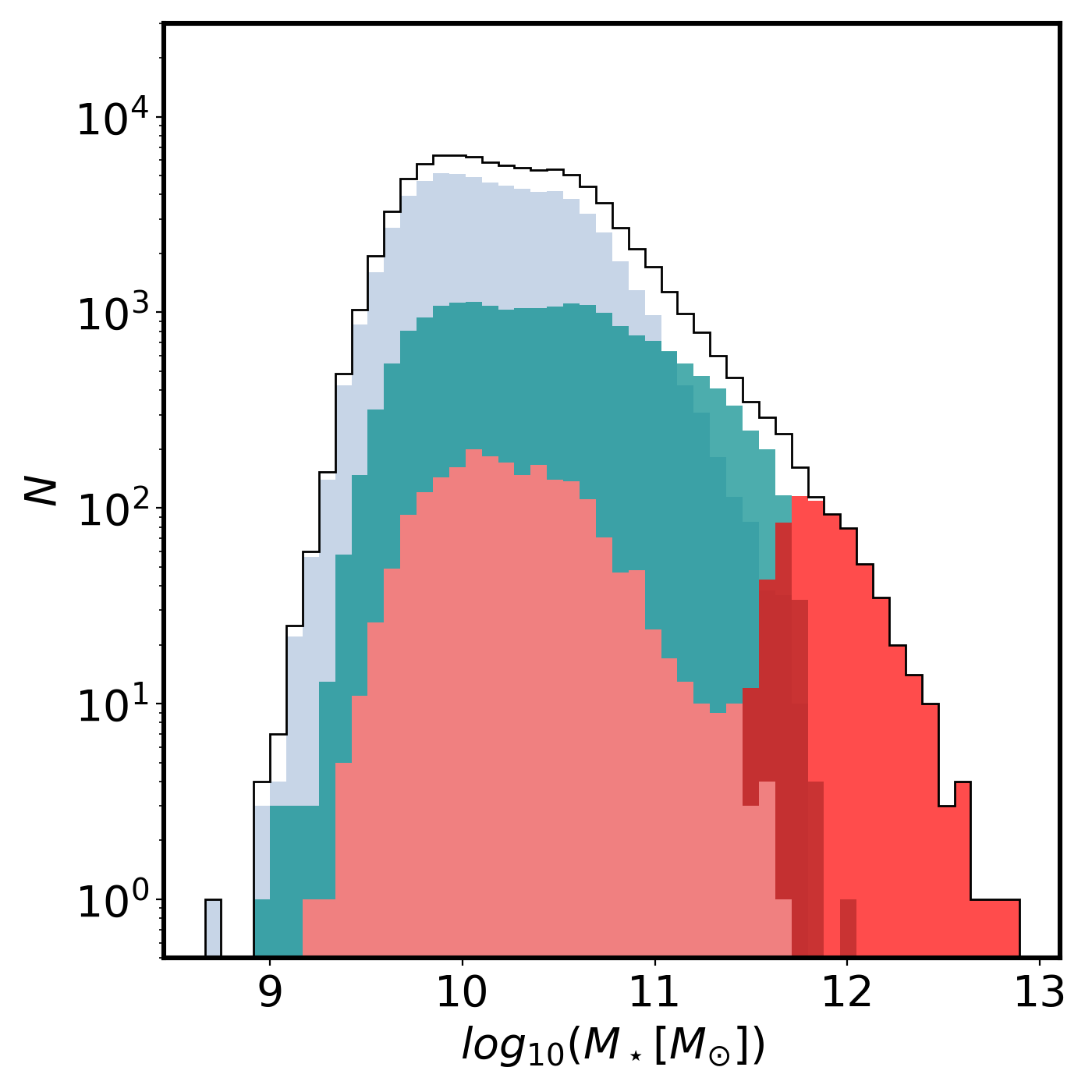}

    \caption{Mass distributions of halos and their central galaxies in TNG300. The top panel shows the distribution of total halo masses ($M_h$), while the bottom panel displays the stellar mass ($M_\star$) distribution of central galaxies. In both panels, the total sample is represented by a solid black line, while different environmental subsets are indicated as follows: \textit{filaments} are shown in green, \textit{clusters} in red, \textit{cluster outskirts} in pink, and \textit{others} in light blue.}
    
    \label{Fig1}%
  \end{figure}

\begin{figure*}
  \centering
 \includegraphics[width=1.99\columnwidth]{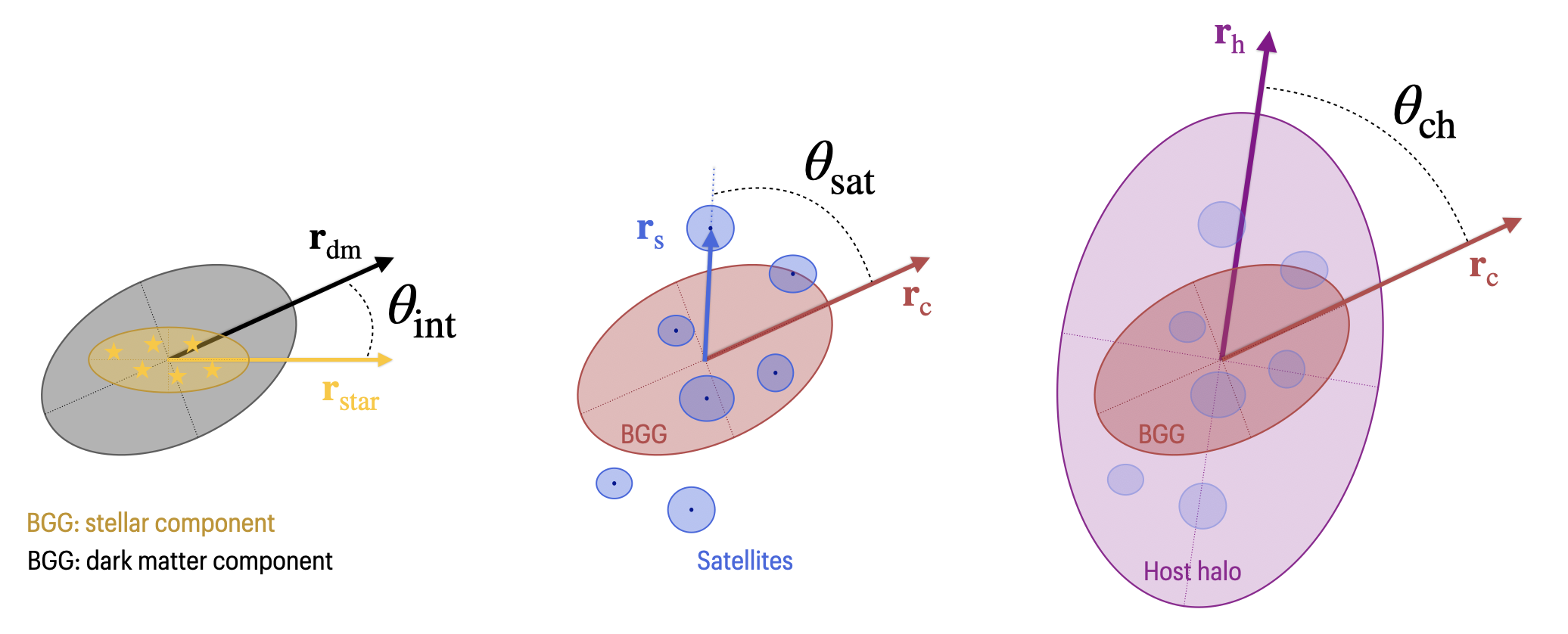}
   \caption{Schematic showing the angles used in this work to study alignments.
From left to right, the angles $\theta_{int}$, $\theta_{sat}$, and $\theta_{ch}$ are displayed. The first, $\theta_{int}$, represents the alignment between the stellar and dark matter components of the BGG. The second, $\theta_{sat}$, corresponds to the angle between the principal axes of the BGG and the position of a satellite galaxy. Finally, $\theta_{ch}$ describes the angle between the principal axes of the BGG and those of the host dark matter halo of the group.}
   \label{Fig0}%
\end{figure*}
To verify the consistency of the simulation results with those observed in large survey detections, we utilized the main galaxy sample from the Sloan Digital Sky Survey Data Release 18 \citep[SDSS DR18,][]{almeida2023}. This release covers an extensive sky area exceeding 10,000 square degrees across five optical bandpasses (u, g, r, i, z) and includes more than 1.2 million galaxies with spectroscopic redshift data, reaching an approximate redshift of ( z $\simeq$ 0.3 ).The spectroscopic version of this survey is statistically complete up to an apparent magnitude limit in the r band of 17.77, which guarantees a robust sample suitable for the analysis of large-scale structures. Each galaxy provides a wealth of information, including its positions, magnitudes, ellipticity, and the position angle of its major axis; the latter two parameters are crucial for the alignment analyses we intend to conduct. These parameters are derived from models based on either Vaucouleurs or exponential profiles.  We have checked that regardless of the model used, the results presented in this work remain consistent. Therefore, following previous studies, this paper will present only the results obtained from the exponential model, pointing out that the figures would be nearly identical if the Vaucouleurs model were used instead.

To identify galaxy groups in the SDSS DR18, we follow the procedure described by \cite{rodriguez2020}. This algorithm combines the Friends-of-Friends (FoF) method \citep{merchan2005} and halo-based techniques \citep{Yang2005}. The algorithm begins by detecting gravitationally bound systems through a percolation method adapted from \cite{Huchra1982}. Each group is assumed to contain at least one bright central galaxy, and its initial properties are estimated based on the total luminosity of its members. In the subsequent iterative step, the halo-based approach refines group memberships and recalculates halo properties on the basis of updated luminosity estimates. This process continues until convergence is reached, ensuring a reliable identification of systems across a broad range of group sizes, from small associations to rich clusters. 
The resulting group catalogue provides essential properties, including galaxy membership, spatial positions, and halo mass estimations ($M_{\text{group}}$), derived using an abundance matching technique \citep{Vale2004,Kravtsov2004,conroy2006,behroozi2010}. This method assumes a one-to-one relationship between a group's characteristic luminosity and its dark-matter halo mass. Additionally, galaxies within each group are classified as central and satellites, with the brightest galaxy designated as the central galaxy, while the remaining members are considered satellites \citep{rodriguez2021}. This catalogue demonstrates strong consistency with weak gravitational lensing mass estimates \citep{gonzalez2021} and effectively reproduces the observed central-satellite galaxy population properties seen in numerical simulations, focusing primarily on the variations in halo occupancy distribution and the properties of member galaxies in different environments \citep[e.g.,][]{alfaro2022, RodriguezMedrano2023}. Our final data set consists of galaxies with spectroscopic magnitudes, redshifts, angular positions, and group and halo properties, providing a robust foundation for analyzing the alignment of galaxies across different cosmic environments.   

\subsection{Large scale environment definitions} 

This work aims to study the alignment of central galaxies located in different components of the cosmic web. We identify the latter using the Discrete Persistent Structures Extractor \citep[DisPerSE;][]{Sousbie2011a,Sousbie2011b} structure finder. DisPerSE \footnote{\url{http://www2.iap.fr/users/sousbie/web/html/indexd41d.html}} detects persistent topological features such as peaks, voids, walls, and particularly filamentary structures using discrete Morse theory \citep[see][]{Morse1934}, operating under the assumption that a mathematical framework, known as the Morse complex, is an adequate description of the cosmic web. In this context, a set of manifolds can be related to the aforementioned cosmic web structures. In a nutshell, this algorithm identifies the critical points of the density field, which correspond to locations where the field’s gradient vanishes—namely, the maximum, minimum, or saddles of the density field. To determine the density field from the input particle distribution (galaxies in the present case), the algorithm employs the Delaunay tessellation field estimator \citep[DTFE; see][]{van2009}.

In this work, we detect filaments in TNG300 following a similar DisPerSE implementation as in \cite{galarraga2020}, namely by using a $3\sigma$ persistence ratio on a smoothed DTFE density field computed from ${\rm M_\ast{}} \geq 10^{9}$~M$_{\odot}$ galaxies. We further associate galaxies to the different cosmic web environments using the following definitions based on distances to cosmic web structures, in a similar fashion to \cite{galarraga2023}:
\begin{itemize}
    \item \textit{Filaments}: These structures correspond to the ridges of the density field between two nodes that connect from node to node. We consider galaxies within 2 Mpc of the filament axis to inhabit this environment.
    \item \textit{Clusters}: This refers to those halos with a virial mass greater than $5 \times 10^{13}$~M$_{\odot}$. We assume that all central galaxies within one virial radius belong to this environment.
    \item \textit{Cluster Outskirts}: These are the regions surrounding the clusters, including those galaxies that lie between 1 and 3 virial radii from the cluster centre.
    \item \textit{Others}: This group includes all central galaxies that are not found in any of the environments defined above. Primarily, they will be those that populate the emptier regions of the universe.
\end{itemize}


The top panel of Figure \ref{Fig1} illustrates the mass distributions of halos (i.e $M_h$ is the mass of the central subhalos provided by the TNG300 simulation), in the different cosmic web environments defined above, specifically those of central galaxies with an r-band brightness greater than -19.5. These are the central galaxies that we will focus on throughout this work. By definition, the halos located in \textit{clusters} occupy the most massive regions. In contrast, the other three environments show a similar mass range, with only minor differences in the average values of their distributions. In terms of the number of objects, the majority of halos are found in the environment categorized as \textit{others}, followed by those in \textit{filaments}, then on the \textit{cluster outskirts} and finally in \textit{clusters}. This distribution pattern fits the volumes occupied by each of these environments, which generally follow the same order. This distribution differs from that found in earlier studies \citep[e.g.,][]{Ganeshaiah2019}, as we only consider bright central galaxies in our analysis. The bottom panel of Figure \ref{Fig1} shows the stellar mass distribution of the central galaxies in the halos, whose (halo) mass distributions were displayed in the top panel. It can be seen that the stellar mass distributions are similar to those shown for the halo mass across the different environments, albeit with greater scatter. 
This reflects a well-established fact: beyond the processes governing halo mass growth, internal astrophysical processes also shape the stellar mass of the central galaxy. This is the reason for the blurring of boundaries that are otherwise clear when considering halo mass alone. In other words, even when imposing a lower or upper halo mass cut to define environments, this does not translate linearly into a selection on the stellar mass of central galaxies. Nevertheless, this does not imply that the significant differences between environments are driven by sampling distinct stellar mass distributions. Therefore, throughout the analysis in this work we will focus on halo mass.

For cosmic web analysis in the SDSS spectroscopic survey, we utilized the identification done by \cite{Malavasi2020a} using the DisPerSE method. This approach followed the same procedure as that used for simulation data while also considering the observational nature of the data and accounting for projection effects. The results of this implementation are publicly available and the derived filament catalogues have been utilized in several studies, including those by \cite{Malavasi2020b}, \cite{Tanimura2020a}, \cite{Bonjean2020} and \cite{Tanimura2020b}.
In this work, we use the $3\sigma$ persistence catalogue based on the Legacy North SDSS galaxy distribution (with smoothed density) consistent with what we employ in TNG300.

To define large-scale environments in SDSS DR18 in a manner analogous to those used in simulations and compare our results, we combine the group catalogue obtained using the method of \cite{rodriguez20} with the DisPerSE implementation of \cite{Malavasi2020a}. We selected central galaxies brighter than -19.5 in the r-band. This approach allowed us to maintain consistent definitions as before: \textit{clusters} were identified using the same mass threshold based on the mass provided by the group identification method; the remaining environments were defined following the same guidelines. 
It is important to note that this environment identification in redshift space introduces systematic effects like the Fingers-of-God elongation and line-of-sight projection artefacts. These effects may dominate the uncertainty budget in environment classification, particularly for cluster regions and filament connectivity.

\section{Alignment definitions}

\label{sec:alignments}
\subsection{Simulation}

In this work, we investigate the alignment properties of central galaxies, focusing on the brightest/most massive group galaxies (BGGs). Building upon our previous methodological developments \citep{Rodriguez2023, Rodriguez2024} and following established best practices for shape measurement \citep{Zemp2011, Bassett2019}, we analyze these systems using exclusively stellar and dark matter particles enclosed within twice the half-dark matter mass radius. We compute the inertia tensor ($I_{i,j}$) of the BGGs as:

\begin{equation}
    I_{i,j} = \sum_n m_n x_i^n x_j^n,
\end{equation} 
where $i$ and $j$ correspond to the three spatial axes of the simulated box, i.e., \(i, j = 1, 2, 3\). The term \(m_n\) represents the mass of the \(n\)-th, while \(x_{i}^n\) and \(x_{j}^n\) denote the positions of the \(n\)-th particle along the \(i\)-th and \(j\)-th axes, respectively. These particle positions are measured relative to the centre of the subhalo to which the particle belongs, defined as the position of the particle with the minimum gravitational potential energy.
Let $\mathbf{r}_{a}$, $\mathbf{r}_{b}$, and $\mathbf{r}_{c}$ be the three normalized eigenvectors of $I_{i,j}$ corresponding to the major, intermediate, and minor axes, respectively.

From these directions, we will define three angles that will provide insight into the internal alignment of the BGG and its alignment with the other group members, referred to as satellites.

We begin by introducing the first angle, which quantifies the alignment of the internal components of the BGG. In this approach, we use the angle between the principal shape axes corresponding to dark matter and stars, which we refer to as $\theta_{int}$, is illustrated in the first diagram of Figure~\ref{Fig0} and is defined as:
\begin{equation}
     \cos(\theta_{int, k}) = {\bf r}_{dm, k} \cdot   {\bf r}_{stars, k}
\end{equation}
where $\theta_{int, k}$ is the angle between the $k$-th semi-axis  calculated using dark matter (${\bf r}_{dm, k}$) and the corresponding semi-axis determined using stars (${\bf r}_{stars, k}$).  

The second angle we define is \(\theta_{sat}\), which is used to analyze the distribution of galaxies in relation to a specific axis of the BGG (see the second diagram of Figure~\ref{Fig0}). This angle is calculated using the following equation:
\begin{equation}
    \cos(\theta_{sat, k}) = \frac{{\bf r}_{dm/stars,k} \cdot  ( {\bf x}_{BGG}-{\bf x}_{sat})}{|{\bf x}_{BGG}-{\bf x}_{sat}|}
\end{equation}
In this equation, \({\bf x}_{BGG}\) represents the position of the BGG, while \({\bf x}_{sat}\) refers to the position of a satellite galaxy. The term \({\bf r}_{dm/stars,k}\) denotes the k-th eigenvector of the BGG's shape, which can be calculated using either dark matter or stars. Essentially, this angle describes the orientation of the satellite's relative position with respect to the BGG's principal axes.

To further explore the alignments and to complement the two previous definitions, we introduce a third angle, denoted as $\theta_{ch}$, and is illustrated in the third diagram of Figure~\ref{Fig0}. This angle is defined as the angle between the principal axes of inertia, which are calculated using the positions and masses of the dark matter particles in the group, and the axes corresponding to the BGG:
\begin{equation}
    \cos(\theta_{ch, k}) = {\bf r}_{c, k} \cdot   {\bf r}_{h, k}
\end{equation}
where $k$ denotes the semi-axis considered and $\bf r$$_c$ and $\bf r$$ _h$ represent the normalized eigenvectors corresponding to the BGG and the group, respectively. 

As we are going to study the behaviour of these three angles statistically throughout this work, it is important to bear in mind that, in the case of a random spatial distribution, they give an average value of 60 degrees. That is, values greater than 60 degrees indicate a misalignment, while those less than 60 degrees suggest an alignment.

\subsection{SDSS}
\label{angulossdss}
To measure the alignments in the SDSS, we introduce angles that are analogous to those defined in simulations while considering the projection limitations of the catalogues. 

The direction of the main shape axis of the BGG is represented by the position angle calculated from its brightness distribution, as provided by the SDSS. The direction of the group is defined by the position angle that corresponds to the eigenvector of the semi-major axis, which is obtained from the two-dimensional inertia tensor using the projected positions of its member galaxies. From this information, the projected counterpart of $\theta_{ch}$ is called $\varphi_{ch}$, calculated simply as the difference between the two angles. 

A similar approach is applied to the analogous angle \( \theta_{sat} \), which we denote as \( \varphi_{sat} \). In this case, we consider the position angle of the projected radius vector of the member galaxies in relation to the main shape axis of the BGG provided by the survey.

It is worth mentioning that, in this case, since these angles are measured in projection, the average values corresponding to a random distribution are 45 degrees.
Moreover, due to the lack of information regarding the dark matter distribution in the BGG halo, we cannot observe an equivalent of \( \theta_{int} \).
In this article, we will first focus on the simulation, reserving the analysis corresponding to the SDSS for the final section.

\section{Dependence of alignments on mass}
\label{sec:mass}

\begin{figure*}
  \centering
  \includegraphics[width=0.67\columnwidth]{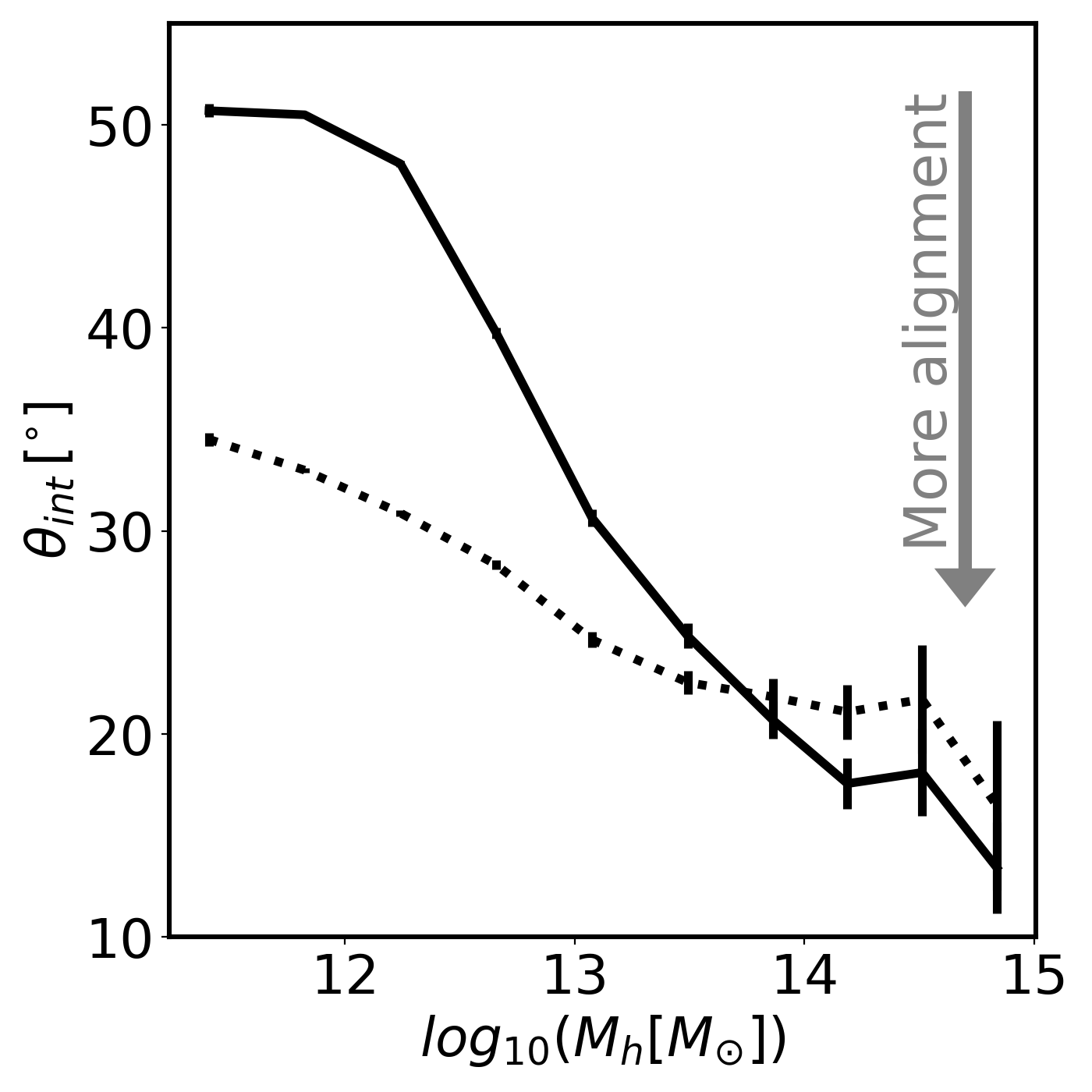}
  \includegraphics[width=0.67\columnwidth]{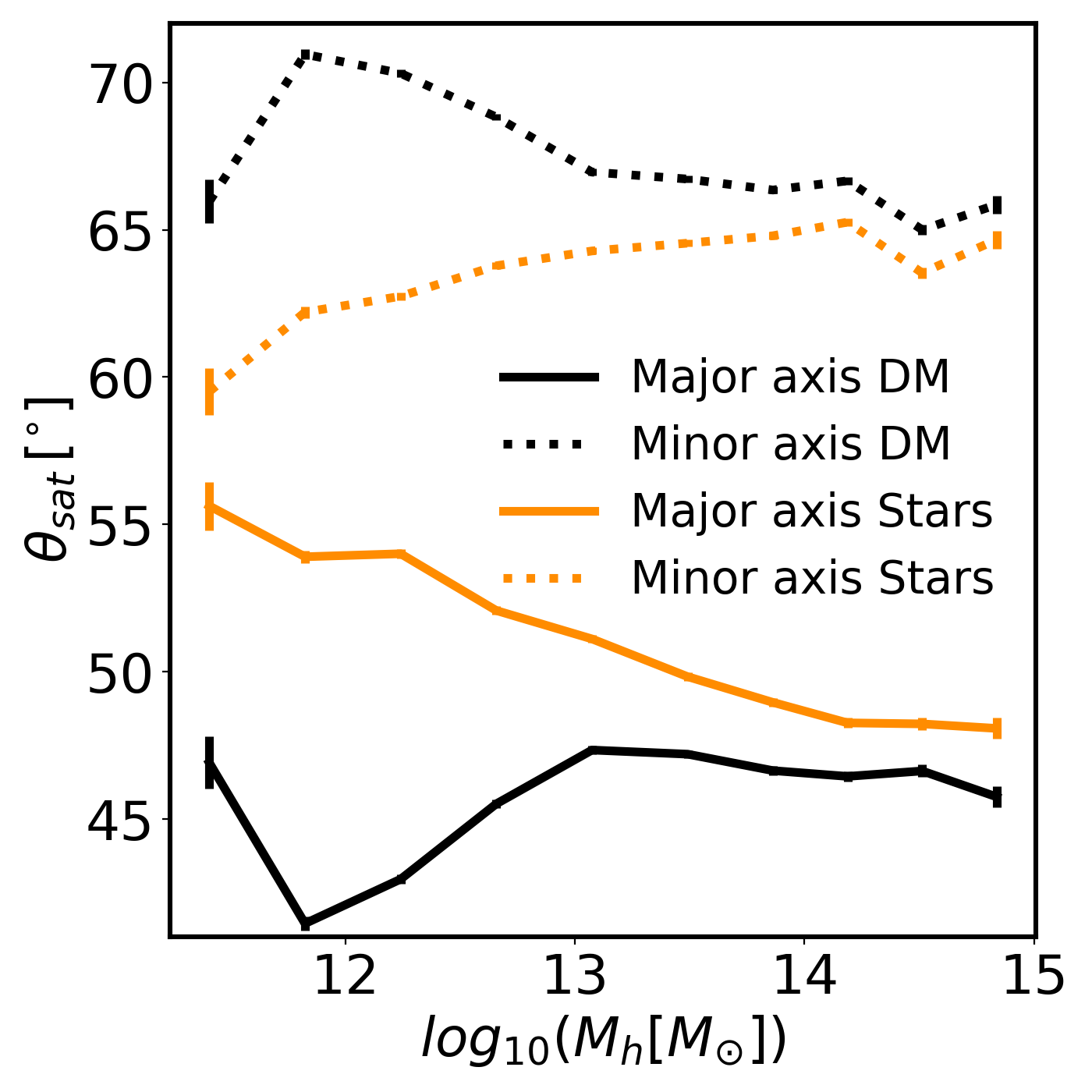}
  \includegraphics[width=0.67\columnwidth]{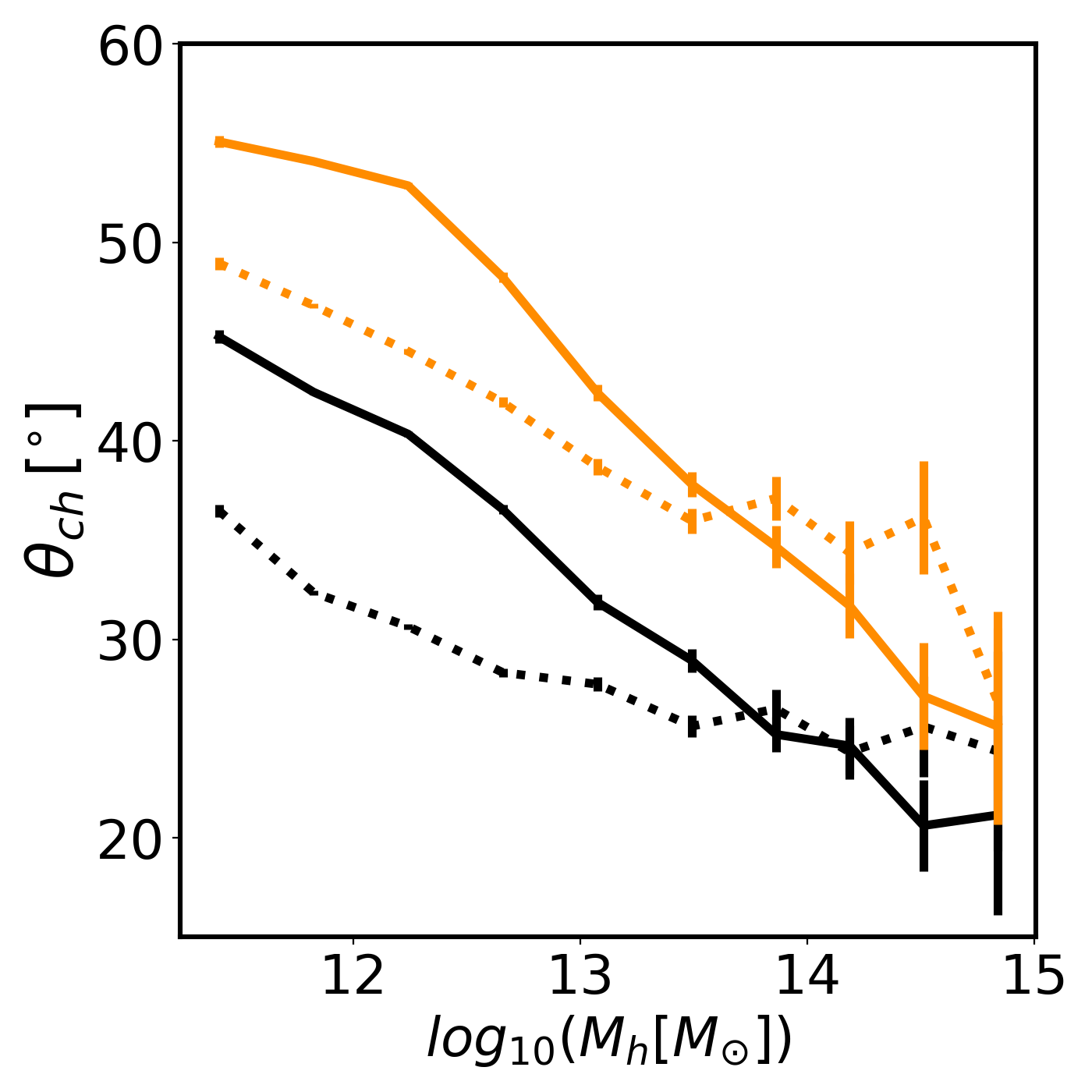}
   \caption{The relation between the alignment and the mass of the halo in which the central galaxies of the sample are located. The left, centre, and right panels present this relation for angles \( \theta_{\text{int}} \), \( \theta_{\text{sat}} \), and \( \theta_{\text{ch}} \) respectively. In all panels, solid lines represent the major axis, while dotted lines represent the minor axis. Additionally, where possible, measurements based solely on stellar information are indicated in yellow. Error bars are calculated using the standard deviation of the mean. The left panel illustrates the direction of increasing alignment. This feature is consistent across all panels displaying alignments in this figure and the subsequent ones.}
   \label{Fig2}%
\end{figure*}

\begin{figure*}
  \centering
  \includegraphics[width=0.67\columnwidth]{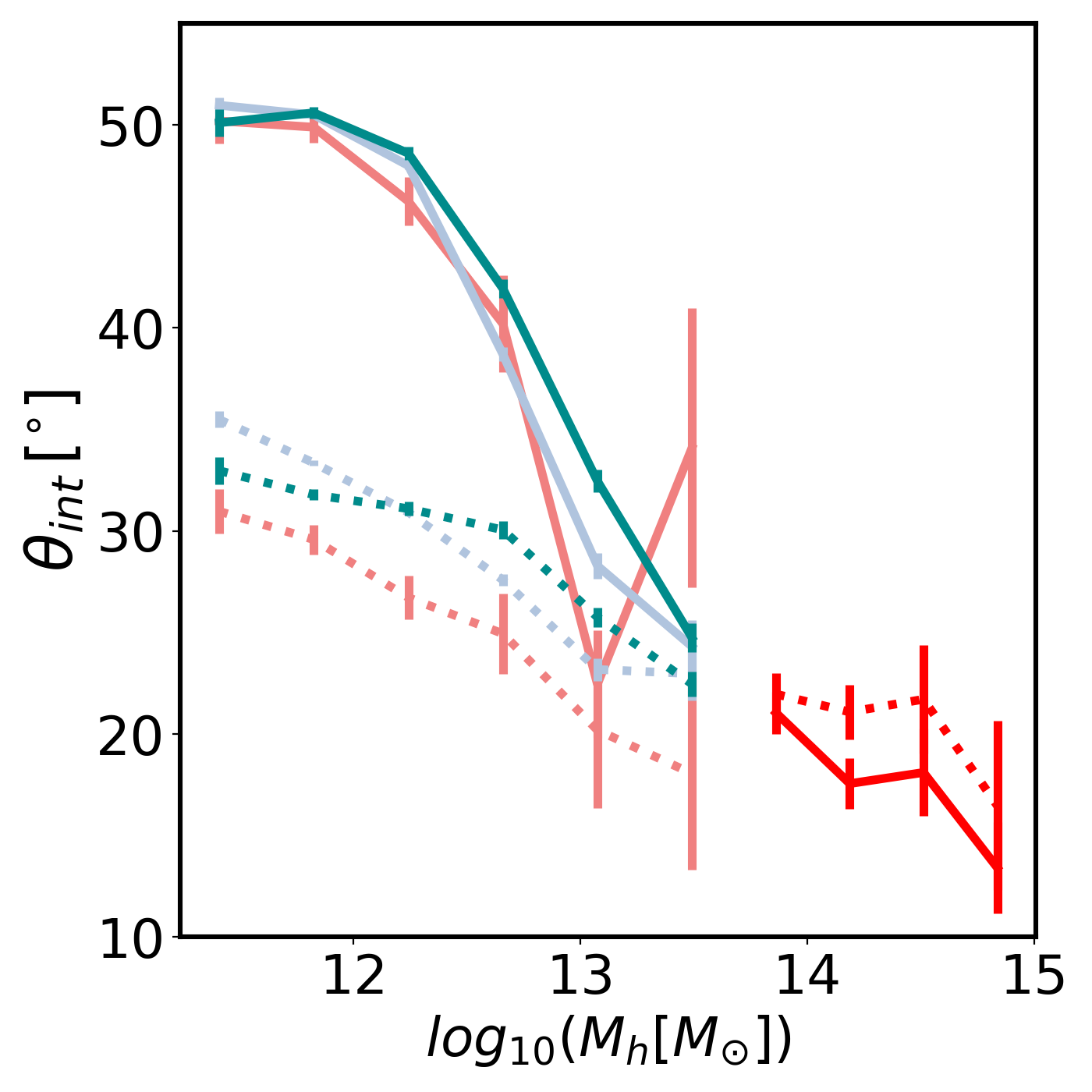}
  \includegraphics[width=0.67\columnwidth]{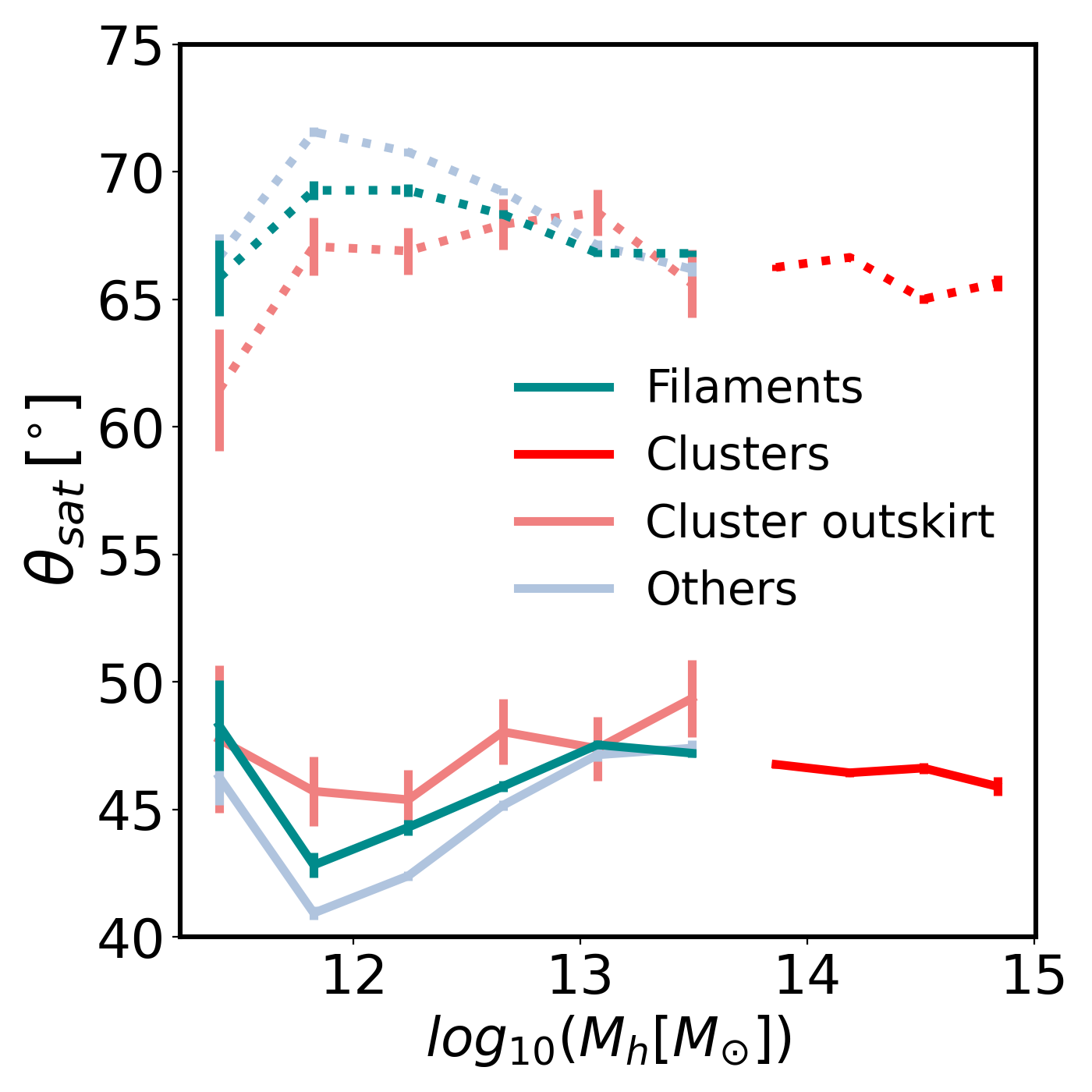}
  \includegraphics[width=0.67\columnwidth]{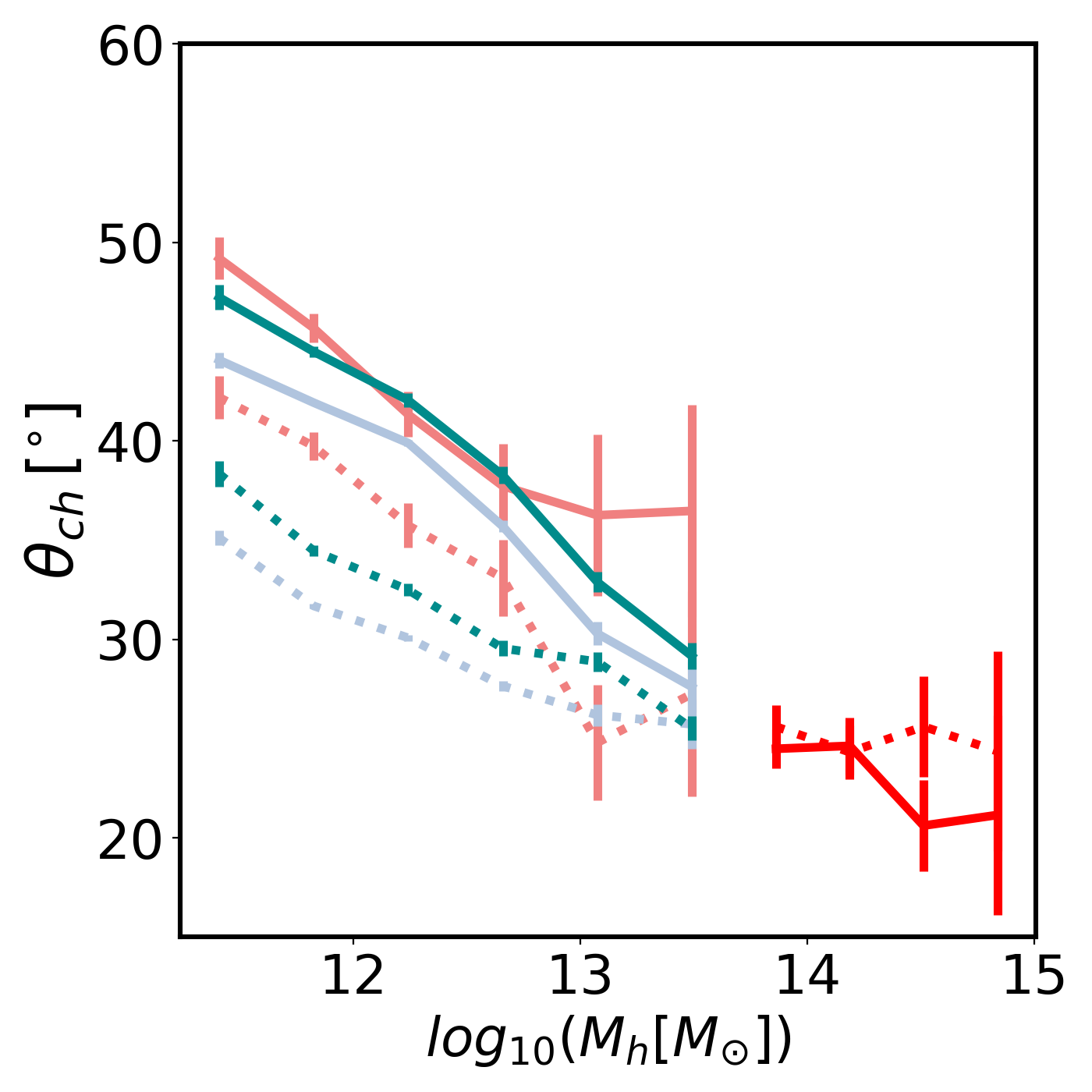}
   \caption{The relation between the alignment of central galaxies and the mass of the halos in which they are located across different large-scale environments. The left, centre, and right panels illustrate this relationship for angles \( \theta_{\text{int}} \), \( \theta_{\text{sat}} \), and \( \theta_{\text{ch}} \), respectively. In all panels, solid lines indicate the major axis, while dotted lines represent the minor axis. The lines shown in green, red, pink, and light blue correspond to halos in \textit{filaments}, \textit{clusters}, \textit{cluster outskirts}, and others environments, respectively. The axes are measured using the dark matter particles and error bars are calculated using the standard deviation of the mean. In all panels, solid lines represent the major axis, while dotted lines represent the minor axis.}
  \label{Fig3}%
\end{figure*}

\begin{figure}
  \centering
  \includegraphics[width=0.8\columnwidth]{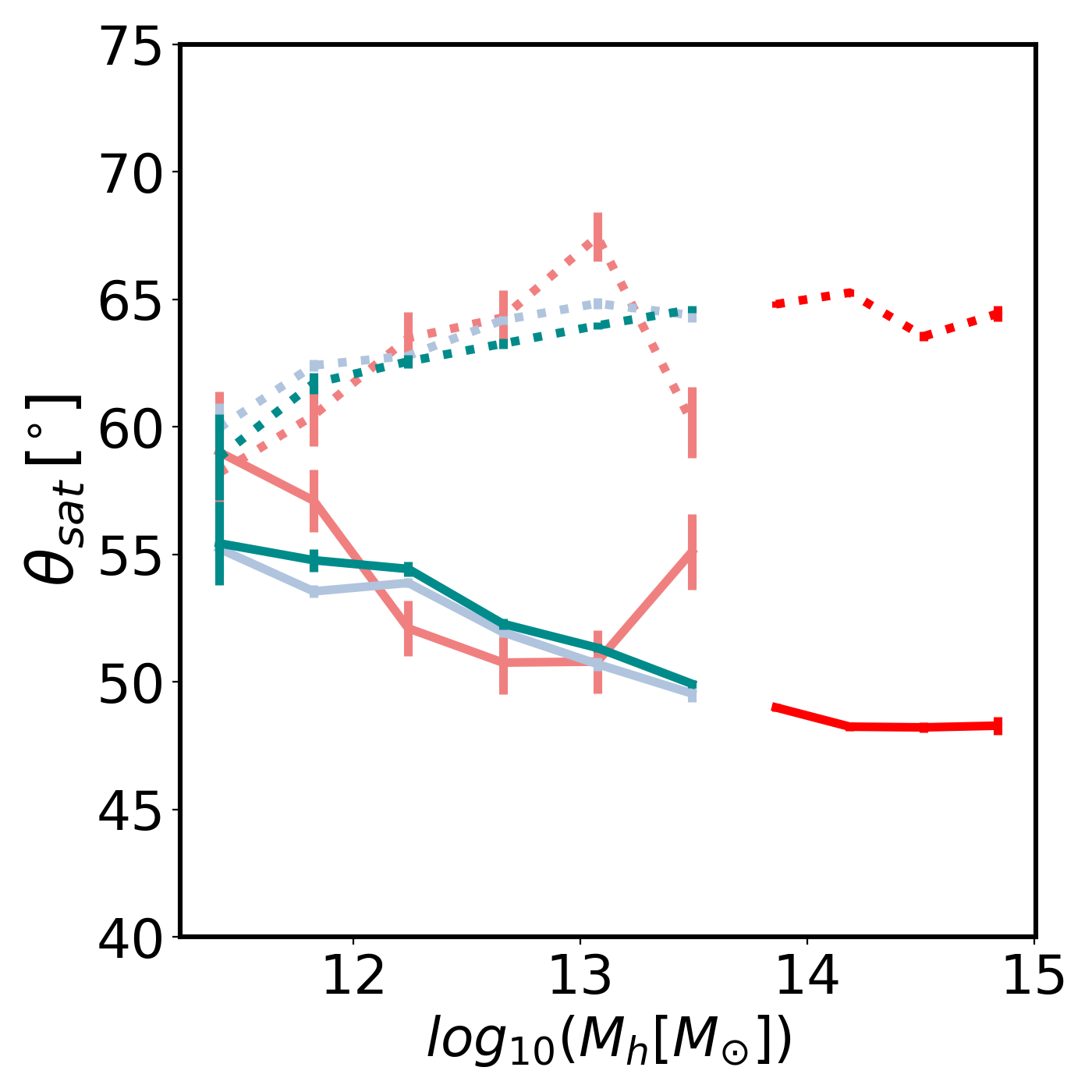}
  \includegraphics[width=0.8\columnwidth]{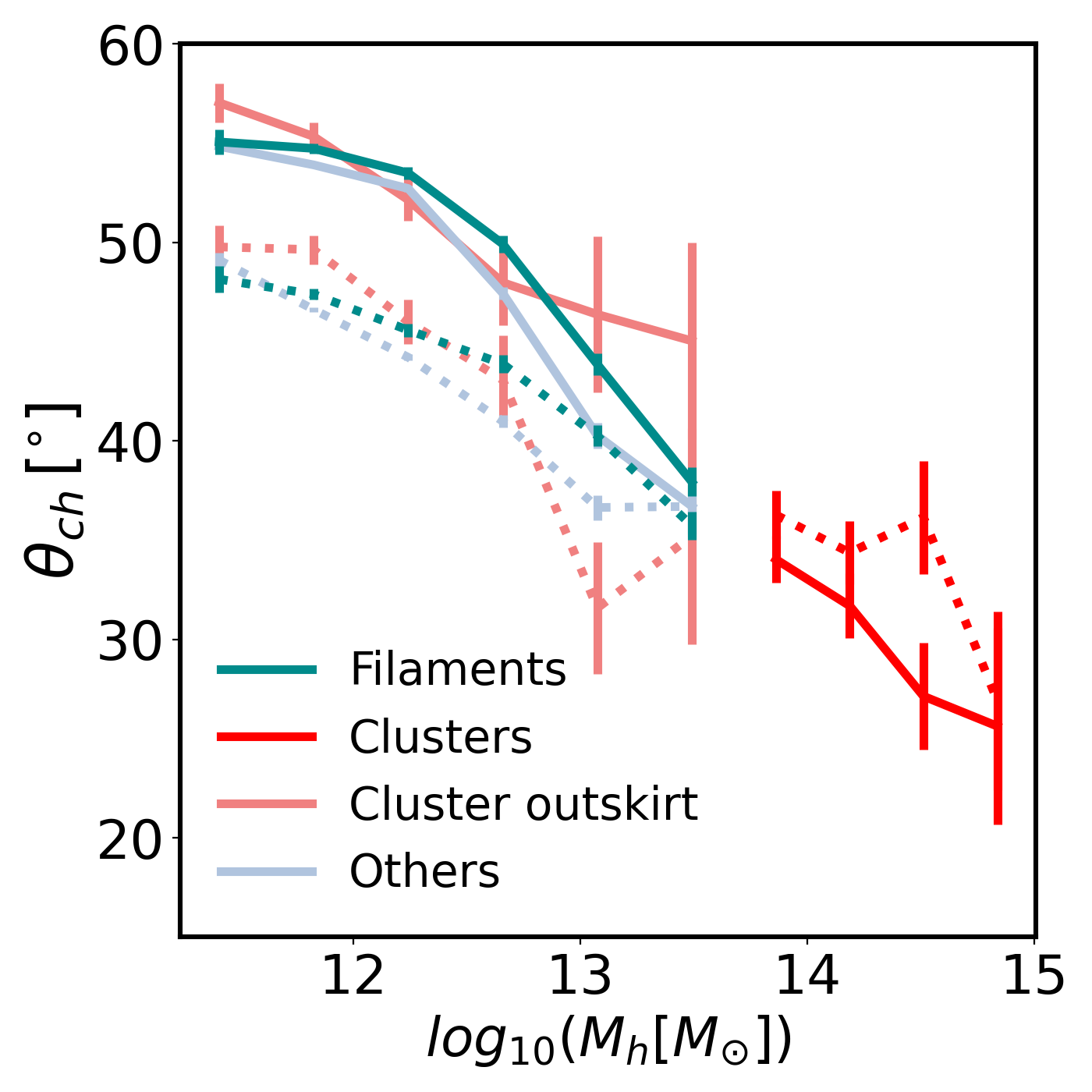}
   \caption{In the same format as in Figure \ref{Fig3}, the alignment of the main axes measured with the stellar component of the central galaxies is presented.}
   \label{Fig4}%
\end{figure}

In this study, we aim to analyze the behavior of the angles discussed earlier with respect to the local and global environments surrounding central galaxies. The mass of a galaxy group serves as the primary indicator of this environment. This mass is closely linked to the astrophysical processes that influence the formation and evolution of the galaxies within the group. For instance, characteristics such as the merger rate, mean density, and colour of the central galaxy show a strong dependence, if not the strongest, on mass \citep[e.g.,][]{Shankar2006, Contreras2015, Wechsler2018, Man2019}. This relationship is also evident in several scaling relations \citep[e.g.,][]{rodriguez2021, behroozi2010}.

In \cite{Rodriguez2022,Rodriguez2023,Rodriguez2024}, we explored the alignments of galaxies, focusing specifically on their colour dependence to compare our findings with the results obtained from observations and simulations. However, those studies showed that the dependence on mass of the alignments calculated using the correlation function is equally strong, if not stronger, than that on colour. Therefore, the first analysis we chose to conduct in the present paper examines the relationship between the alignment of the central galaxies in our sample and the mass of the halo in which they are located.

In the left panel of Figure \ref{Fig2}, we observe that the angle \( \theta_{int} \) decreases as the mass increases for both the major and the minor axes. The minor axis exhibits values lower than the major axis for masses below \(\sim 10^{14} M_{\odot} \), while both axes show similar values for masses above this threshold. This angle represents the alignment between the stars and the dark matter halo they inhabit, indicating that the major and minor axes of both the stars and the dark matter halo tend to become more aligned as the mass of the halo increases. Additionally, it suggests that for lower mass haloes, the minor axes are more closely aligned with each other than their respective major axes. The noticeable difference in alignment between the major and minor axes of galaxies residing in less massive halos may be attributed to their tendency to be spiral (oblate) \citep[see, for example,][]{Rodriguez2024}. This morphology results in a more clearly defined minor axis compared to the other two axes, which are quite similar and lie within the same plane. In the most massive end, the predominant morphology is triaxial, meaning that the three axes are clearly defined, with no differences in alignment between the minor and major axes. This behavior of internal alignment may arise from the fact that the processes involved in the formation and evolution of galaxies vary depending on their mass. This variation underscores the complex nature of galaxy development, highlighting how halo mass plays a crucial role in shaping galaxies.

The central panel of Figure \ref{Fig2} illustrates the relationship between $\theta_{sat}$ and mass. It is evident that this angle, measured with respect to the major (resp. minor) axis, shows a strong alignment (resp. misalignment) across the entire range of masses. 
Examining the dark matter estimates (black curves), we observe that unlike many other halo properties, $\theta_{sat}$ does not show a strong dependency on mass. Only for lower masses, around \(\sim 10^{12} M_{\odot} \), alignment in both axes seems to be a small feature. 
It's important to note that this angle is calculated for each satellite individually. As a result, at the lower halo mass end, there are many groups with only a few members. In contrast, at the higher mass end, there are fewer clusters but a greater number of satellites.
Estimating the axes of the central galaxy using stellar material yields measurements closer to 60 degrees. Here, the alignment of the major axis with mass increased, but the minor axis showed the opposite trend. This behaviour is expected since, as illustrated in the left panel described above, the stars are misaligned in relation to their own dark matter halo, and this misalignment decreases as halo mass increases. As a result, measurements derived from dark matter and stars tend to converge and resemble each other as halo mass increases. The results of $\theta_{sat}$ indicate that satellite galaxies are preferentially distributed around the major axis.

We finally focus on the angle between the central galaxy's shape with that of the host group, \(\theta_{ch}\), in the right panel of Fig. \ref{Fig2}. Is it clear that this angle significantly decreases with mass, showing a progressive alignment between the shape of the galaxy and its associated group as halo mass increases. This trend is observed for both the major and minor axes when analyzing the shapes of dark matter and stellar components. Notably, measurements taken from stars exhibit greater misalignment than those related to dark matter. This discrepancy may result from the internal misalignment represented by \(\theta_{int}\), indicating that the dark matter of the BGG is more closely aligned with the shape of the host halo than the stars of the BGG \citep[see figures 7 and 8 in][]{Rodriguez2023}.

Previous studies have shown that central galaxies are aligned with their surroundings, and this alignment is more pronounced when considering dark matter instead of just stars. In this section, we analyzed how these alignments depend on halo mass and found that both the internal orientation angle \((\theta_{int}\)) and the orientation angle of the central galaxy shape with respect to that of the host halo (\(\theta_{ch}\)) increase with mass. In contrast, the distribution of satellites (\(\theta_{sat}\)) does not exhibit a significant dependence on mass.

\section{Environmental effects}
\label{sec:environment}

\begin{figure*}
  \centering
  \includegraphics[width=0.67\columnwidth]{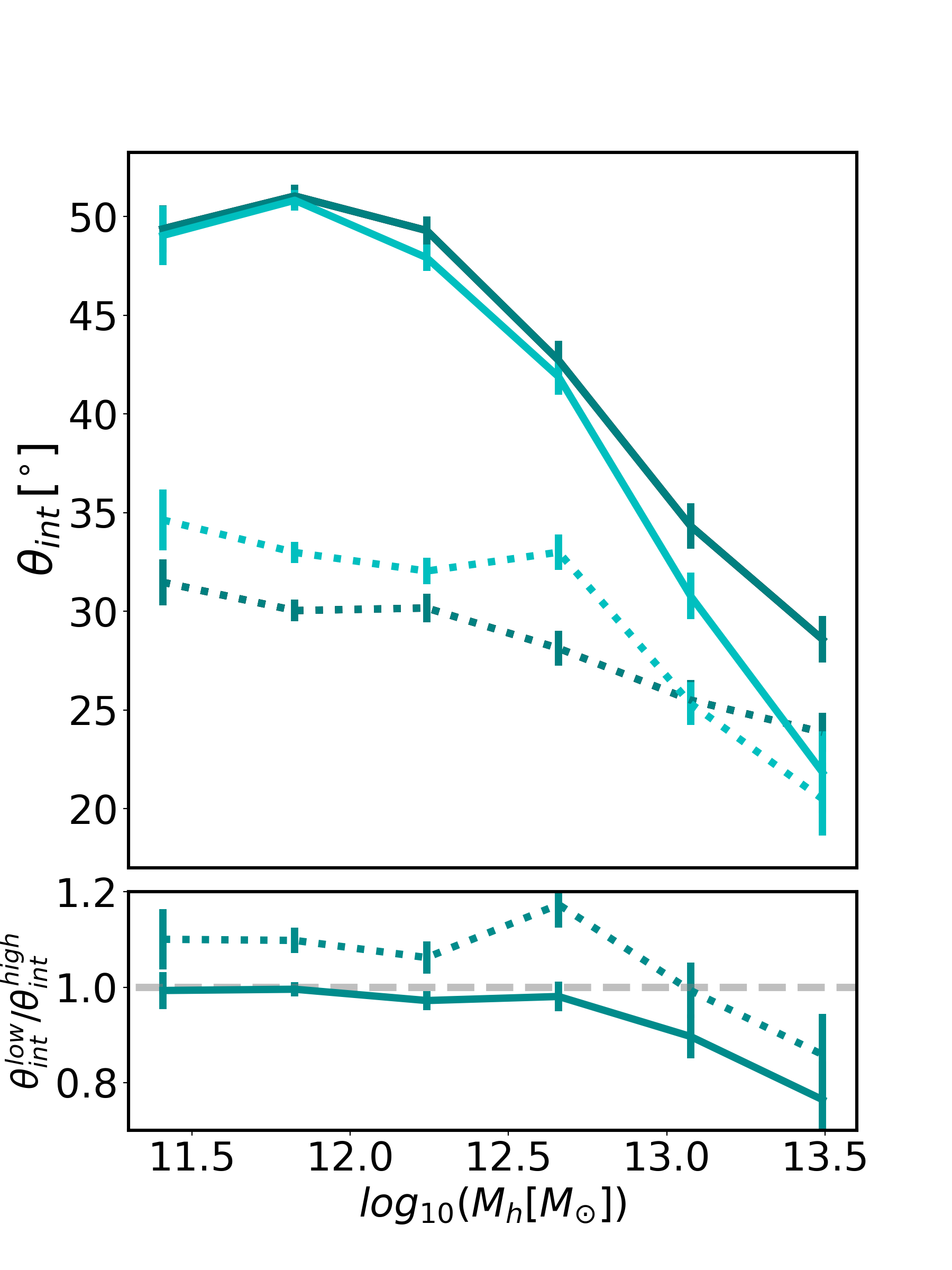}
  \includegraphics[width=0.67\columnwidth]{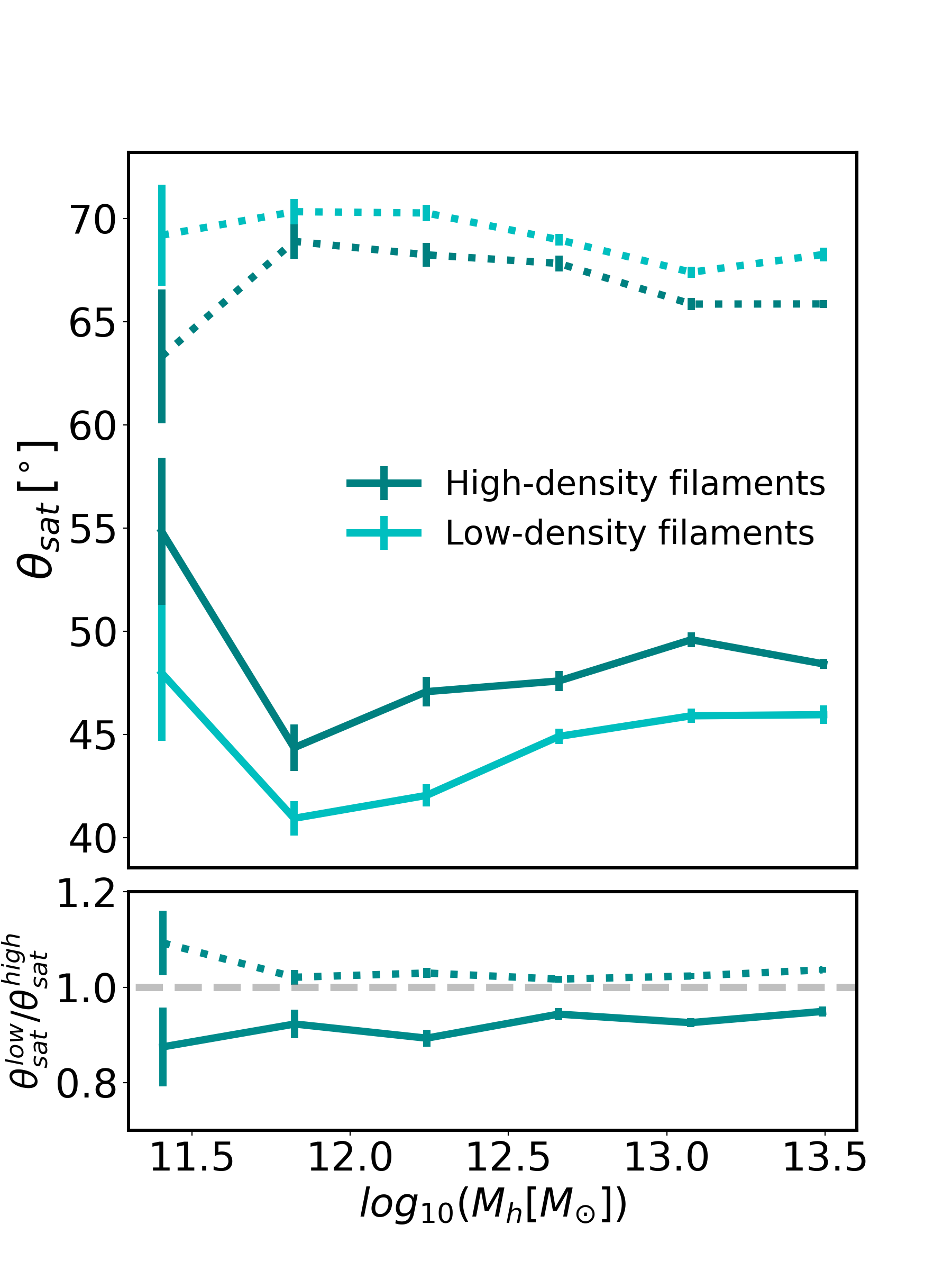}
    \includegraphics[width=0.67\columnwidth]{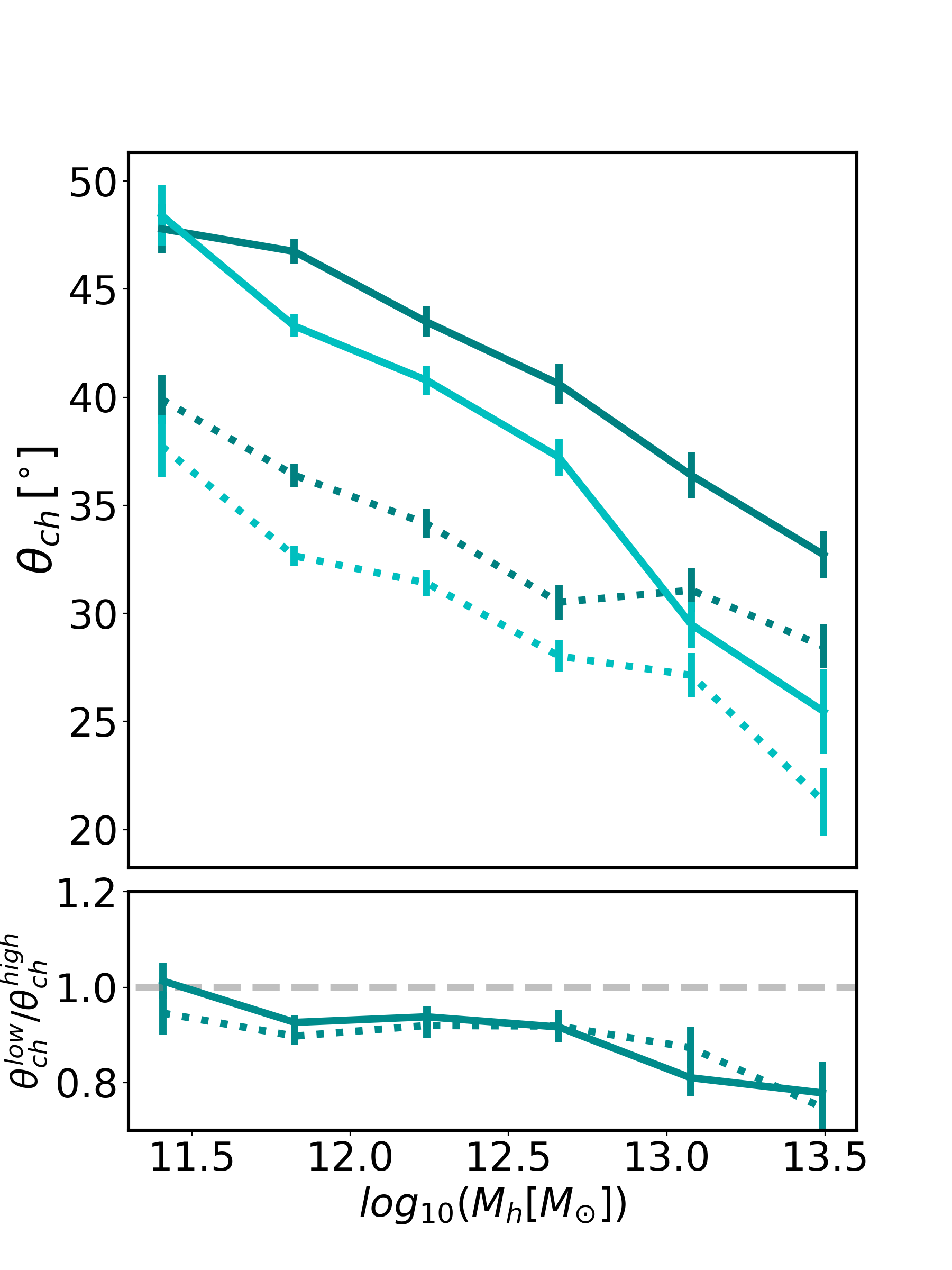}
   \caption{Dependence on halo mass the angles $\theta_{int}$ (left), $\theta_{sat}$ (centre), and $\theta_{ch}$ (right) for central galaxies located in \textit{filaments} of varying densities. Darker lines represent the highest density quartile, while lighter lines correspond to the lowest. In all panels, solid lines represent the major axis, while dotted lines represent the minor axis. The lower panels show the ratio between the results at different densities.}
   \label{Fig5}%
\end{figure*}

\begin{figure}
  \centering
  \includegraphics[width=0.9\columnwidth]{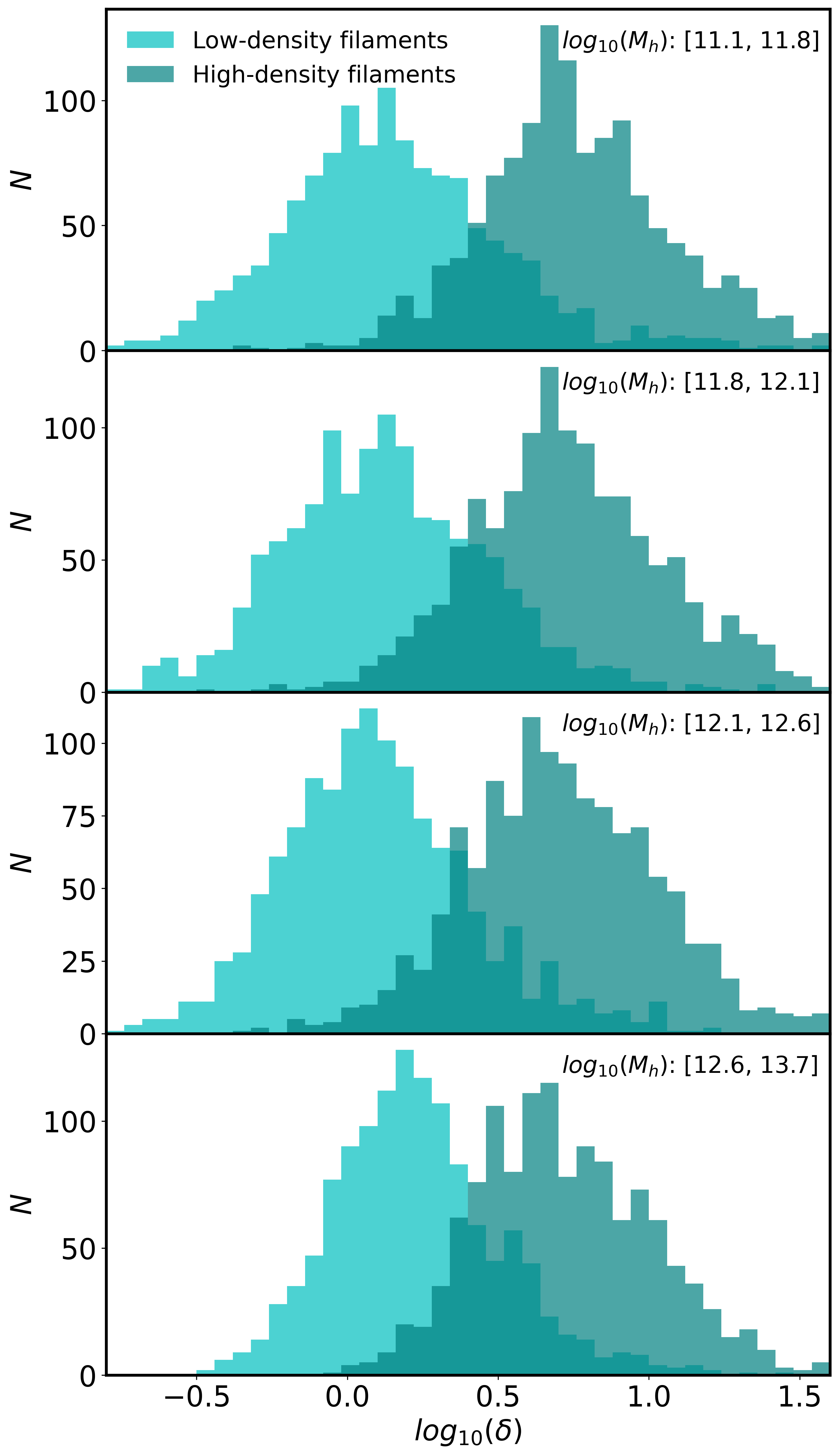}
   \caption{Local density distribution at 5 Mpc for central galaxies found in low- and high-density filaments. }
  \label{Fig5bis}%
\end{figure}

\begin{figure}
  \centering
  \includegraphics[width=0.9\columnwidth]{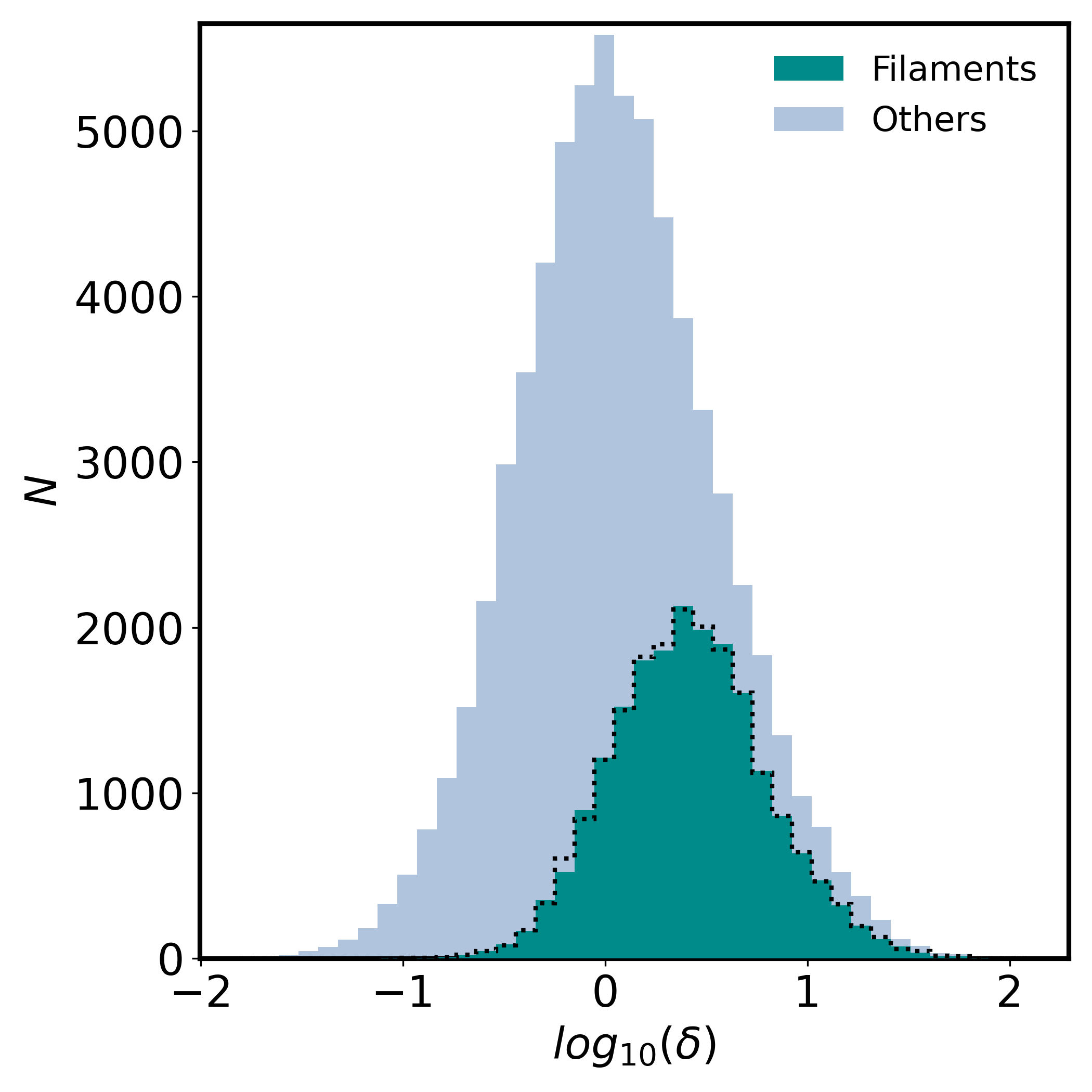}
   \caption{Local density distribution at 5 Mpc for central galaxies found in \textit{filaments} and those classified as \textit{others}. The dotted line represents the distribution of galaxies selected from the \textit{others} that correspond to those in \textit{filaments}.}
     \label{Fig6}%
\end{figure}

\begin{figure*}
  \centering
  \includegraphics[width=0.67\columnwidth]{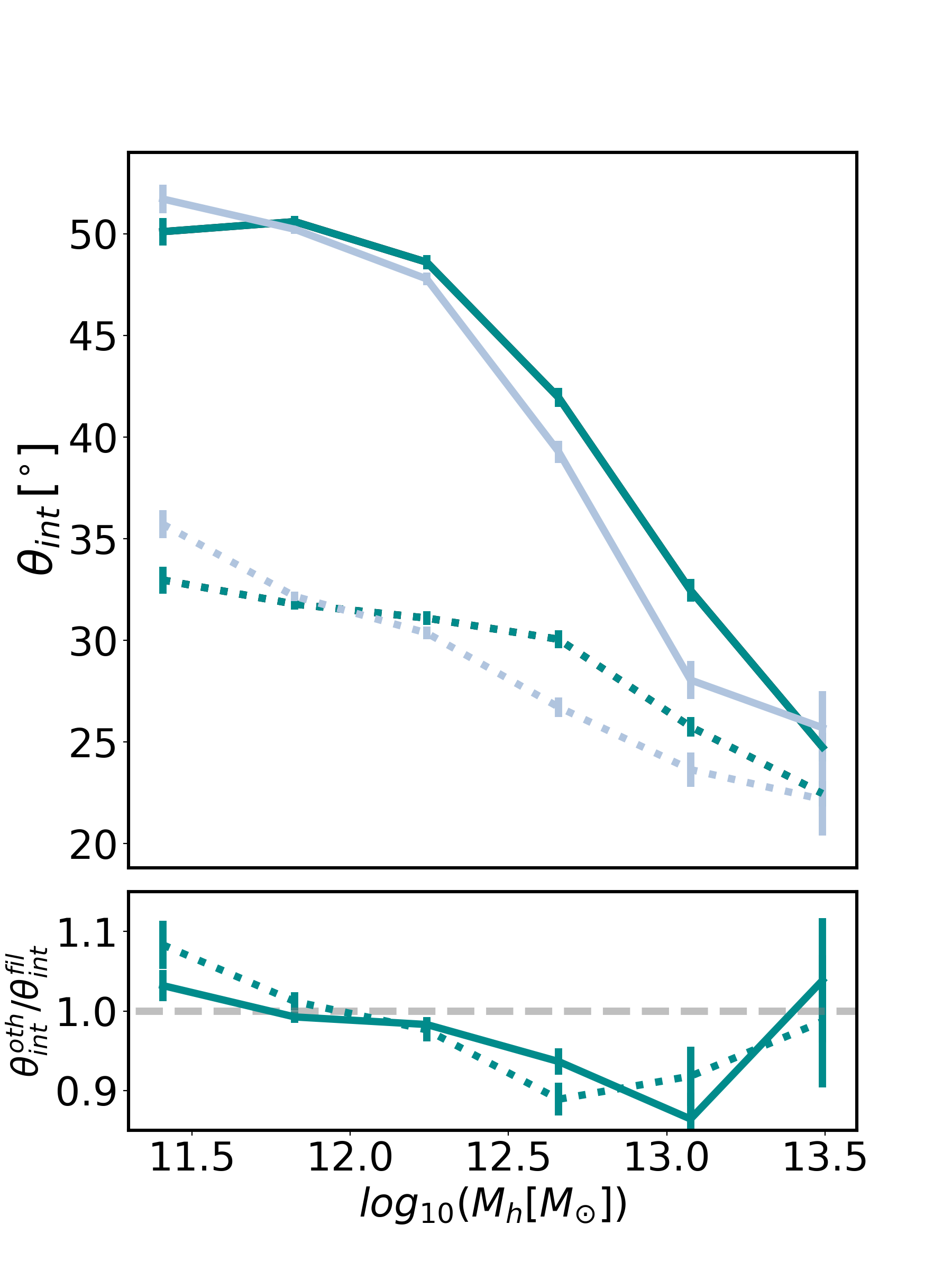}
  \includegraphics[width=0.67\columnwidth]{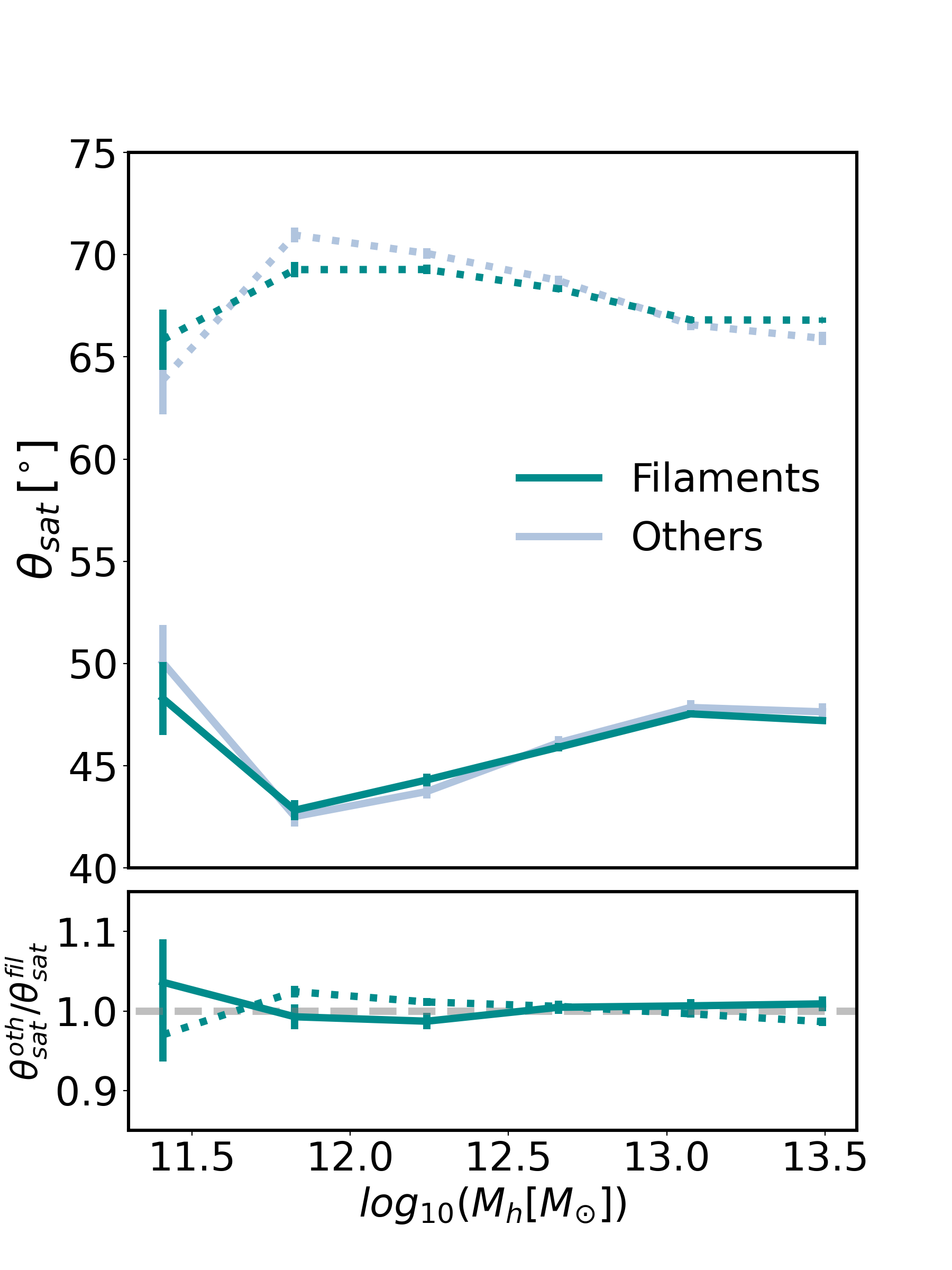}
    \includegraphics[width=0.67\columnwidth]{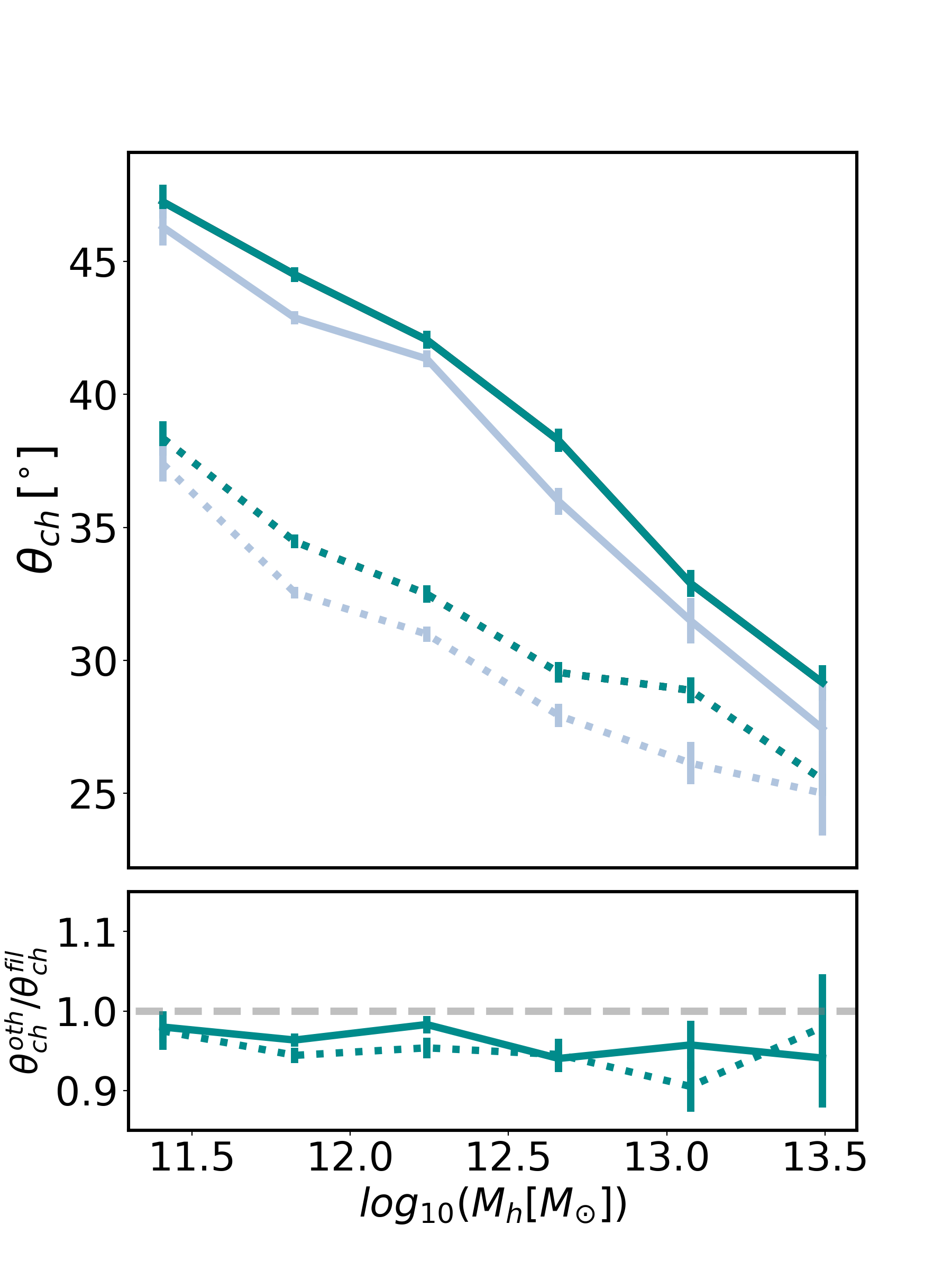}

   \caption{Using the same format as Figure \ref{Fig5}, results are shown for central galaxies in \textit{filaments} or \textit{others} with the same local density distribution.}
  \label{Fig7}%
\end{figure*}

Having established the relationship between the studied angles and halo mass, we now will analyze how the large-scale environment and its characteristics influence these alignments in \textit{filaments}, \textit{clusters}, \textit{cluster outskirts}, and \textit{others}. Given that clusters occupy uniquely the high mass-end of the halo mass distribution, the trends for galaxies in these structures will be dominated by the mass effect described above. Regardless, we will include this information for the sake of completeness.  In contrast, the remaining environments are populated by objects within a similar and overlapping halo mass range, as illustrated in Figure \ref{Fig1}, thus enabling the exploration of any specific trends arising from environmental effects while keeping the halo mass fixed. To begin with, we analyze the alignments taking into account the axes defined using dark matter particles. This builds on the results derived from the black curves presented in the Figure \ref{Fig2}.

In the left panel of Figure \ref{Fig3}, we illustrate how \(\theta_{int}\) varies as a function of halo mass across different environments. At a glance, we can see that in all environments the same trend with halo mass is observed as in the overall sample of the previous section/figure. Along the major axis (solid lines), we observe that around \( 10^{13} \, M_{\odot}\) stellar material and dark matter exhibit lower alignment in \textit{filaments} than \textit{others}, as the $\theta_{int}$ values for galaxies in \textit{filaments} are $\sim \, 4^\circ$ higher than in \textit{others}. Note that the \textit{cluster outskirts} shows noisy results, mainly because there are only 2122 halos in this environment.  In contrast, galaxies in \textit{others} demonstrate the lowest degree of alignment at lower masses. Additionally, within the halo mass range from approximately \(10^{12}\) to \(10^{13} \, M_{\odot}\), central galaxies in \textit{filaments} are less aligned compared to those in other environments. Although the above results are statistically significant, the differences found are minimal across the different environments compared to the trends with mass, which are the dominant ones driving the alignments. 

In the middle panel of Figure \ref{Fig3}, we observe a variation in the alignment of \(\theta_{sat}\) with cosmic environment. Galaxies classified as \textit{others} show a somewhat stronger alignment along the major axis, followed by those located in \textit{filaments}, while galaxies on the \textit{cluster outskirts} exhibit a weaker alignment. Conversely, the results for the minor axis present an opposite trend. With this in mind, the results suggest that the alignment of satellite galaxies increases as the density decreases.

For the different environments, the right panel of Figure \ref{Fig3} shows the relationship between \( \theta_ {ch} \) and mass. For the minor axis (dotted lines), it is observed that there is a lower alignment for the centrals in \textit{cluster outskirts}, followed in alignment by the \textit{filaments} and, finally, by those in \textit{others}. On the major axis, the trends are similar although the results for \textit{cluster outskirts} are noisy. This would indicate that, like \(\theta_{sat}\), the dark matter halos of central galaxies are increasingly aligned with the host halos, when the environment is less and less hostile.\\

We now analyse the results for \( \theta_{sat} \) and \( \theta_{ch} \) using stellar particles. They are shown in Figure \ref{Fig4}. In general, we observe that in both panels, the differences obtained between environments are smaller than the corresponding results for the case where the axes are defined using the dark matter distribution. For \( \theta_{sat} \), it is not possible to detect any variation of alignment with environment given the uncertainties in the measurements. In the case of \( \theta_{ch} \) (lower panel), although there appear to be some differences similar to those observed with dark matter. However, it should be borne in mind that, to compare it with observations, we would need to consider the effects of projection, the distortions produced by the redshift space, the errors inherent in the measurements or the environment detection, which together contribute to an increase in the errors of the angles we are studying.

\subsection{Variations with Filament density}

The previous results suggest a second order connection between large-scale environments and the alignment of central galaxies. We will now further explore this connection by addressing potential differences within the \textit{filaments} class, namely by disentangling any variation of alignment as a function of the density of the filaments hosting central galaxies.
To do this, we define filament density based on the galaxies located within 2 Mpc of the filament's axis, considering a cylinder of the same radius, with length corresponding to the total length of the segments that comprise the filament.

In Figure \ref{Fig5}, we analyze the behaviour of the angles \(\theta_{int}\), \(\theta_{sat}\), and \(\theta_{ch}\) by dividing the \textit{filaments} based on their density. Two samples were taken using the first and fourth quartiles of the filament densities, corresponding to 24.5 and  82.7  $Mpc^{-3}$, respectively. Note that, to calculate this density, we considered all galaxies with stellar mass greater than $10^{8.5} M_{\odot}$. Darker lines represent higher-density filaments, while lighter lines indicate lower-density ones.
In the left panel, we examine the alignment between dark matter and stars ($\theta_{int}$). For denser filaments, the stellar component is less aligned with its dark matter halo along the major axis, particularly at higher masses. In contrast, the minor axes exhibit the opposite trend, with greater alignment in denser environments. This is clearly shown in the lower panel, which quantifies the ratio of angles between low- and high-density filaments, $\theta^{low} / \theta^{high}$. For the major axis (solid line), we observe differences of up to 20\% at high masses, while the minor axis (dashed line) shows smaller differences of around 10\% at low masses.
The middle and right panels present \(\theta_{sat}\) and \(\theta_{ch}\), respectively.  Both panels show that for both minor and major axes, there is a greater alignment in less dense \textit{filaments}. In summary, the only exception to this trend is the alignment \(\theta_{int}\) for the minor axis at lower masses, which behaves differently than the other alignments.
Regarding the ratio of angles between low- and high-density filaments for \(\theta_{sat}\), our analysis demonstrates that only the major axis is responsive to filament density, showing consistent differences of approximately 10\% throughout the halo mass range. Furthermore, the ratio for \(\theta_{ch}\) remains similar for both axes and increases with mass, spanning from very low values to greater than 20\%. These results underscore the strong effect of different filament environments on the alignment of central galaxies with respect to the three different reference scales explored in this paper.

The results indicate that in the outermost regions of the halo traced by $\theta_{sat}$ and $\theta_{ch}$, the denser \textit{filaments} generate a misalignment with respect to the major and minor axes of the central galaxy. On the other hand, in the innermost regions (traced by $\theta_{int}$), dark matter and stars tend to align with the minor axis of the galaxy.

\subsection{Variations with local density}

At this stage, a key question emerges: are the filaments (as physical structures) directly influencing the alignments, or are we simply measuring local density effects? In other words, are the \textit{filaments} or the local environment responsible for affecting the alignments? 
Therefore, it is important to examine how much these samples differ in local density. 

To address this issue, we estimate local density by calculating the density of dark matter within a sphere of radius 5 Mpc centred around each BGG. We compare the local density distributions of central galaxies found in low- and high-density filaments, as illustrated in Figure \ref{Fig5bis}. We present the results across four halo mass ranges to avoid potential biases associated with the mass variability. Due to the significant differences in local density observed between the two samples, we found that adjusting for local density resulted in alignment differences that closely resembled those shown in Figure \ref{Fig5}, which classifies galaxies according to filament density (we do not display this figure here because it does not provide relevant information).

We take a step further and focus now on the differences between centrals in \textit{filaments} and those in \textit{others}, i.e. the most under-dense cosmic environments. Figure \ref{Fig6} illustrates the local density distribution for those galaxies classified as belonging to \textit{filaments} and \textit{others}. As can be expected, the distributions differ significantly, with the filaments being typically much denser. In order to determine whether the presence of filaments has any effect on the alignments beyond local density effects, we make the two samples comparable by selecting a subsample of BGGs from the others that match the density distribution of the filaments sample. This is illustrated by the dotted lines in Figure \ref{Fig6}. 

Following the same approach as in the previous sections, Figure \ref{Fig7} shows the results of the alignments for the BGGs in filaments (darker lines) and the mimic sample of other (lighter lines). In the innermost regions of the halo, when analysing $\theta_{int}$, we observe that, at an equal local density, the differences between the field and the filaments are equivalent to those in Figure \ref{Fig2}, where the local density was not matched. This suggests that the observed differences are indeed due to the presence of the filament. As for the distribution of satellites around the BGG, represented by $\theta_{sat}$, the central panel clearly illustrates that there are no significant differences when comparing the filament categories with the others, both with equal distribution of local densities. However, when analysing $\theta_{ch}$, it is observed that, at equal local density, the presence of the filament reduces the alignment on both axes, although the differences are not as marked as in Figure \ref{Fig2}. This suggests that the filament influences the alignment of the dark matter halo but not so much the distribution of satellite galaxies, or at least not to a level detectable with the sensitivity of our experiment.  

\section{SDSS}
\label{sec:sdss}

\begin{figure*}
  \centering
    \includegraphics[width=1.99\columnwidth]{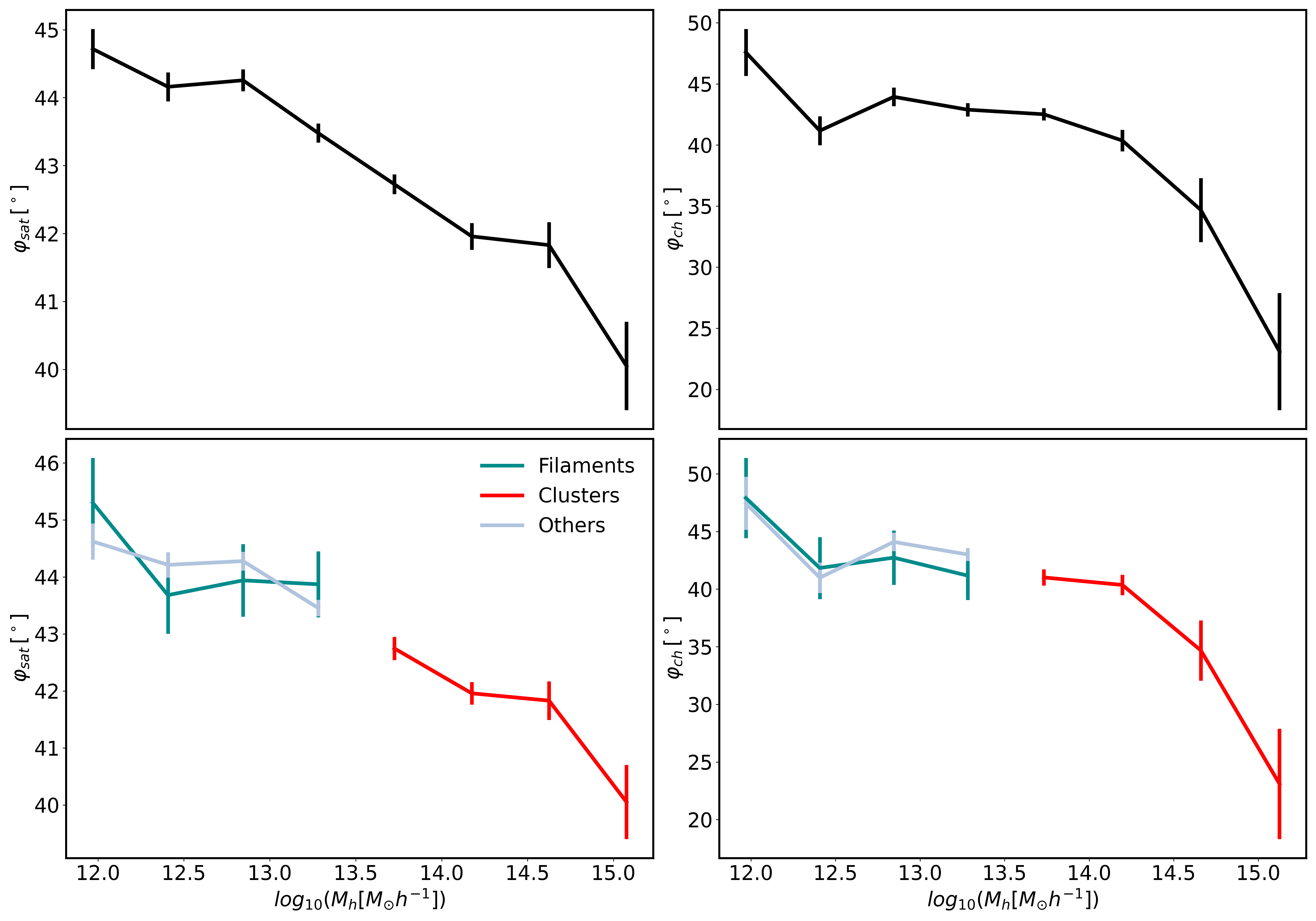}

   \caption{SDSS measurements of the alignments of the major axis of the central galaxies with the location of the satellite galaxies ($\varphi_{sat}$) and of the major axis of the central galaxies with the major axis of the group to which they belong, measured with the member galaxies ($\varphi_{ch}$). The upper panels show the total sample and its dependence on mass, while the lower panels show those corresponding to different environments of the cosmic web. In all panels, error bars represent the standard deviation of the mean.}
  \label{Fig8}%
\end{figure*}

Numerical simulations are important tools because they allow the analysis of complex phenomena and enable highly accurate predictions. One of their key advantages is that they are not influenced by observational biases that can affect studies relying solely on observed data. However, it is essential to compare theoretical predictions with actual observations to validate the results and draw reliable conclusions.
In this section, we will analyze the alignments of central galaxies in the Sloan Digital Sky Survey (SDSS). To do this, we will use the data and the angles defined in Sections \ref{datasdss} and \ref{angulossdss}, respectively.

The first step is to measure how the alignments depend on mass. In this context, we will examine the projected versions of \(\theta_{sat}\) and \(\theta_{ch}\), which are denoted as \(\varphi_{sat}\) and \(\varphi_{ch}\). Since the data we are using does not provide information about the dark matter content of the BGG, it is not possible to obtain the internal alignment.
To calculate the shape of the group in the catalogue we only have the member galaxies provided by our identifier. Generally, the number of members in these groups is relatively low; it is uncommon to find groups with more than ten members. Therefore, to maximize our sample size and obtain a reasonable estimate of the main axes, we select groups with at least four members for our shape calculations, which we will use to study \(\varphi_{ch}\).

The results are shown in the upper panels of Figure \ref{Fig8}, where we only present those corresponding to the major axis of the BGG. As these angles are two-dimensional measurements, the results regarding the minor axis are complementary.

We first note that, despite the projection, we recover the trends detected in the simulations, as the alignment of the satellites \( \varphi_{sat} \) (shown in the upper left panel) shows a strong dependence on the mass. Similar to the simulations, no alignment is observed in low-mass systems; however, alignment significantly increases as halo mass increases (for reference, see measurements indicated by the yellow lines in Figure \ref{Fig2}). It is remarkable to find this dependence in the observations and that it agrees with the simulations. Indeed this result supports the reliability of the abundance matching procedure, which assigns masses in the group catalogue based solely on luminosity ranking and aims to provide a proxy for the mass, rather than an exact measurement.

In the upper right panel of figure \ref{Fig8}, we show the projected alignment of the BGG shape with respect to the halo shape, $\varphi_{ch}$. Once again, the behaviour of $\varphi_{ch}$ as a function of halo mass resembles that of the simulations: for low masses, the alignment is undetectable, whereas for high masses, it increases. It should be emphasised that, besides the uncertainties in determining the mass, the shapes are estimated with very few members and, therefore, the direction of the major axis of the group has significant uncertainty, unlike the simulations where the shape is determined with greater precision using numerous dark matter particles.

Based on our observations, we have provided observational confirmation of an important relationship between galaxy alignments and their mass. The next logical step is to investigate how these alignments depend on their environmental conditions on a large scale. For this analysis, we will focus exclusively on three types of environments: \textit{filaments}, \textit{clusters}, and \textit{others}. We are omitting \textit{cluster outskirts} due to the limited number of galaxies in that category, which prevents us from obtaining reliable results. Bottom panels of Figure \ref{Fig8} show the corresponding results for each environment.

As observed, it is not possible to differentiate the alignments between the trends of \textit{filaments} and \textit{others} environments. This is not surprising, given several predictable factors. First, simulations indicate that stars are misaligned with their halo (see Figure~\ref{Fig5}), which causes an attenuation in the signal we detect between the dark matter halo of the BGG and the group’s halo. Second, the identification of environments is less precise compared to simulations due to known observational biases. Considering these factors, which influence how galaxies are assigned to different environments, we assessed the impact of projected distance cuts on our results by testing several definitions of filament thickness. However, the results remained consistent with those presented in Figure \ref{Fig8}, convincing us to keep the original definitions.

\section{Summary and conclusions}
\label{sec:conclusions}



This study builds on our previous findings \citep{Rodriguez2022,Rodriguez2023} to explore how the mass of host halos and the locations of galaxies, whether in \textit{clusters}, \textit{filaments}, \textit{cluster outskirts}, or \textit{other} environments, affect these alignments.

We analyse how three different angles characterise the alignment of central galaxies with their multi-scale environment to understand their relation to the structure on different scales. First, we study the alignment between the stellar and dark matter components of the central galaxy ($\theta _{int}$), which provides information on the internal alignment. Second, we examine the angle between the satellite galaxies and the shape axes of the central galaxy ($\theta _{sat}$), allowing us to characterise the distribution of stellar material within the halo in which the group resides. Finally, we investigate the alignment between the shape axis of the central galaxies and that of the group’s host halo ($\theta _{ch}$), providing information on the connection between galaxy and structure on larger scales These angles are illustrated in Figure \ref{Fig0}, which provides a schematic representation of their definitions. These analyses let us explore how the surroundings and multi-scale physical processes influence the angles characterising the central galaxies.

First, we performed our analysis using the TNG300 hydrodynamical simulation. We explored how each angle depended on the halo mass. We found that both $\theta_{\text{int}}$ and $\theta_{\text{ch}}$—measured using the shape axes of the central galaxy defined by either the dark matter or stellar distributions— exhibit a clear trend of decreasing with halo mass (see Figure \ref{Fig2}). The first one suggests that, internally, the stellar component becomes more tightly coupled with dark matter as the halo mass increases. Similarly, from $\theta_{\text{ch}}$ we can infer that the dark matter and stellar components of the central galaxy show a stronger alignment with the dark matter of the host halo as mass grows. In contrast, $\theta_{\text{sat}}$ does not demonstrate a clear dependence on mass when measured using the dark matter axes. When we consider the axes defined by the stellar distribution, a clear increasing trend with halo mass emerges. 
However, in all cases where we measure the angles using the axes defined by stellar distribution, we find that it is systematically weaker than when measured using the dark matter axes. This suggests that while the stellar structure somewhat follows the dark matter, its degree of alignment with its surroundings is less pronounced. This could mean that the astrophysical processes affecting stars are more complex than those influencing dark matter, which is reflected in their alignments. 
These results expand on previous works, such as \cite{Valenzuela2024}, where the authors study  central galaxy shapes in the Magneticum simulation and conclude that stellar and dark matter shapes are correlated at internal scales, but the inner halo appears to be decoupled from the outer halo. We find that the strength of both the internal stellar–subhalo  and the subhalo–host halo alignment are strongly coupled in the more massive halos, and the large-scale structure environment impacts more on the subhalo–host halo alignment than on the internal alignment. 

Another analysis we conducted was the study of the variation of alignments as a function of halo mass across different environments of the cosmic web (see Figure \ref{Fig3}). We classified these environments into four categories: \textit{clusters}, \textit{filaments}, \textit{cluster outskirts}, and a category labelled \textit{others}, which includes voids and walls. As expected, the most massive central galaxies —also exhibiting the strongest alignments— are by definition in \textit{clusters}, while the other three environments host central galaxies of lower masses.
We found significant differences between environments, suggesting that alignments depend on large-scale structure. Since we compare galaxies at fixed mass, the detected variations reflect the impact of the environment beyond the effect of the host halo mass.  While the mass of the host halo largely influences alignments, the location in the most hostile environments of the cosmic web appears to disrupt these alignments at a fixed halo mass.
Furthermore, since our previous analyses were based on dark matter, we investigated whether these differences persisted when measuring the principal axes of central galaxies using their stellar component (Figure \ref{Fig4}). We observed that the differences decreased compared to those found with dark matter, indicating that environmental influence is more pronounced in the dark matter structure than in the stellar distribution.

To further explore the relationship between alignments and the environment, we specifically analyzed the case of \textit{filaments} and studied the dependence of alignment angles on filament density (see Figure \ref{Fig5}), with the goal of probing the extreme diversity of the filament population. We found systematic variations with density, which suggests that filament properties influence the alignments of central galaxies consistent with the previous analysis, i.e. central galaxies in denser filaments are more misaligned.
This finding led us to investigate whether the local density or the presence of the filaments themselves is the determining factor. As expected, we found a strong correlation between filament and local density (see Fig. \ref{Fig5bis}). This suggests that the effects observed in different environments could be associated with the characteristic local density of each region, rather than the presence of these specific structures. To address this question, we selected central galaxies in the areas labelled as \textit{others} with a local density distribution similar to that of filaments (see Fig. \ref{Fig6}). In this way, any detected differences should be attributed to effects related to their location in different regions of the cosmic web (see Fig. \ref{Fig7}). Thus, we identified that, when analyzing $\theta_{\text{int}}$, there are differences in the alignments between the stellar and dark matter components when central galaxies are located in filaments or field at the same local density. However, the distribution of satellite galaxies, in terms of $\theta_{\text{sat}}$, does not seem to be affected by the presence of these structures. Additionally, the alignment in $\theta_{\text{ch}}$ also shows less coherence when the central galaxy is in filaments compared to the field. The results indicate that filaments cause a misalignment effect on dark matter host halos and the central galaxies within them, particularly affecting their dark matter component. In contrast, the positions of satellite galaxies appear to be less influenced by the presence of these filaments. This analysis showed that, in general, local density is the driving factor in the alignments of central galaxies, and that the position within the large-scale environment plays a secondary, marginal, role.

To contrast the results of the TNG300 simulations, we performed a similar analysis using SDSS data. First, we examine the alignments as a function of halo mass (see Figure \ref{Fig8}) and find a clear dependence, similar to what we observed in the simulations. This finding is particularly significant given the numerous observational biases inherent in SDSS data that could potentially obscure the signal, unlike in simulations where such biases are absent. The consistency between the two analyses underscores the reliability of group identification and halo mass assignments in the observational data. Additionally, it suggests that the physical effects that influence galaxy alignments in the real universe are reflected in the simulations. Secondly, we investigated whether the differences observed between environments in the simulations were also present in the observational data, focusing on three environments: \textit{clusters}, \textit{filaments} and \textit{others}. However, we did not detect significant differences, likely due to observational biases that blur the effects identified in simulations. Nevertheless, it would be interesting to explore this further using alternative filament and void identification methods applied to future observational data.

In conclusion, this study has deepened our understanding of the alignments of central galaxies concerning their environment, both internally and on larger scales, using hydrodynamical simulations and observational data. The results confirm that the halo mass, and environment (both local and, to a lesser extent, large-scale), play a crucial role in determining these alignments. However, open questions remain about how other factors, such as the processes of galaxy formation and evolution, might influence these patterns. 
For example, \cite{Xu2023_TNG300} analyzed the alignment between the principal axes of the stellar component and the host halo as a function of mass, and showed that fixing the ex situ stellar mass fraction almost eliminates the alignment’s dependence on halo mass, meaning that mass accretion  processes strongly influence the alignments.
Also, other studies using the SAMI and MaNGA surveys have established that there exists a relationship between the stellar mass and the alignment of galaxies' angular momentum with filaments, and that some evolutionary processes such as bulge growth, AGN activity and mergers play an important role in producing the so-called spin flip effect \citep{Welker2020,Kraljic2021,Barsanti2022,Barsanti2023,Barsanti2025}. Given that galaxy shape orientation can be related to their stellar angular momentum, there may be a connection between the spin–filament alignment and the axis alignments studied in this work. Nonetheless, caution is needed when making comparisons, as these are different types of alignments. 
Therefore, future research projects could explore in more detail the impact of the accretion history of galaxies or the relationship between galaxy morphology and their alignment in different environments. Additionally, it would be beneficial to validate these findings through larger volume simulations like Flamingo \citep{Schaye2023} or MillenniumTNG \citep{Pakmor2023}, which would enhance statistical analysis. Exploring new methods for identifying structures in observational data would also help refine our understanding of the connection between galaxies and the cosmic web. This work is particularly timely, as new data from upcoming spectroscopic surveys such as DESI (Dark Energy Spectroscopic Instrument), Euclid, and PFS (Prime Focus Spectrograph) will soon be available. These surveys will provide unprecedented statistics and resolution, enabling rigorous tests of our predictions and offering new insights into galaxy alignments within the cosmic web.

\begin{acknowledgements}
      The authors wish to thank the anonymous referee for her/his report that helps us to improve this manuscript. 
      FR, MM and AVMC thanks the support by Agencia Nacional de Promoci\'on Cient\'ifica y Tecnol\'ogica, the Consejo Nacional de Investigaciones Cient\'{\i}ficas y T\'ecnicas (CONICET, Argentina) and the Secretar\'{\i}a de Ciencia y Tecnolog\'{\i}a de la Universidad Nacional de C\'ordoba (SeCyT-UNC, Argentina).
      FR and ADMD thank the ICTP for their hospitality and financial support through the Junior Associates Programme 2023–2028 and Senior Associates Programme 2022–2027, respectively.
      DGE acknowledges Volker Springel for the financial support that facilitated the research visit during which part of this work was conducted. 
      MCA acknowledges support from ANID BASAL project FB210003.
      ADMD thanks Fondecyt for financial support through the Fondecyt Regular 2021 grant 1210612.
\end{acknowledgements}


\bibliographystyle{aa} 
\bibliography{main} 
\end{document}